\documentclass{IEEEtran}
\usepackage{latexsym}
\usepackage{graphicx}
\usepackage{amsfonts,amssymb,amsmath}
\usepackage{hyperref}
\usepackage{cite}
\usepackage{url}
\usepackage{multirow}
\usepackage{algorithmic}
\usepackage{ulem}
\usepackage{balance}
\usepackage{textcomp}
\usepackage{scalerel}
\usepackage{color}
\usepackage{subfigure}
\def\BibTeX{{\rm B\kern-.05em{\sc i\kern-.025em b}\kern-.08em
    T\kern-.1667em\lower.7ex\hbox{E}\kern-.125emX}}
\begin{document}
\title{Hybrid Channel Modeling and Environment Reconstruction for Terahertz Monostatic Sensing}
\author{Yejian Lyu, Zeyu Huang, Stefan Schwarz,~\textit{Senior Member, IEEE}, and Chong Han,~\textit{Senior Member, IEEE} 
\thanks{Y. Lyu and C. Han are with the Terahertz Wireless Communications Laboratory at Shanghai Jiao Tong University, Shanghai 200240, China (e-mail: \{yejian.lyu; chong.han\}@sjtu.edu.cn);

Z. Huang and S. Schwarz are with the Institute of Telecommunications, TU Wien, Vienna, Austria, (e-mail: \{zeyu.huang; stefan.schwarz\}@tuwien.ac.at).
%(Corresponding author: Chong Han).
}
}

\maketitle

\begin{abstract}
Terahertz (THz) integrated sensing and communication (ISAC) aims to integrate novel functionalities, such as positioning and environmental sensing, into communication systems. Accurate channel modeling is crucial for the design and performance evaluation of future ISAC systems. In this paper, a THz measurement campaign for monostatic sensing is presented. Vector network analyzer (VNA)-based channel measurements are conducted in a laboratory scenario, where the transmitter (Tx) and receiver (Rx) are positioned together to mimic monostatic sensing. The centering frequency and measured bandwidth for these measurements are $300$\,GHz and $20$\,GHz, respectively. A directional scanning sounding (DSS) scheme is employed to capture spatial sensing channel profiles. Measurements are conducted across $28$ transceiver (TRx) locations arranged along an `L'-shaped route. Then, an element-wise space-alternating generalized expectation-maximization (SAGE) algorithm is used to estimate the MPC parameters, i.e., amplitude and delay. Specular and diffuse reflections are analyzed based on geometric principles and the estimated MPC parameters, where the effects from the radiation pattern are observed. A geometry-based MPC trajectory tracking algorithm is then proposed to classify the MPCs and de-embed the effects of the radiation pattern. Following this algorithm, a hybrid channel model is proposed based on the de-embedded MPC parameters. In this hybrid channel model for monostatic sensing, the multipath components (MPCs) are categorized into target-related and environment-related components. The target-related components are utilized for target detection and identification, while the environment-related ones focus on geometrical scenario reconstruction. A demonstration of geometrical environment reconstruction, along with an analysis of reflection loss for target identification, is subsequently presented. This work offers valuable insights into THz monostatic sensing channel modeling and the design of future THz ISAC systems.
\end{abstract}


\section{Introduction}
\label{sec:introduction}
\IEEEPARstart{I}{ntegrated} sensing and communication (ISAC) is anticipated to be one of the core technologies in sixth-generation ($6$G) networks, enabling communication systems to not only transmit data but also detect and identify environmental targets~\cite{isac_jhzhang,6G_isac1,6G_isac2,isac_low_freq_model}. ISAC systems for $6$G are expected to deliver millimeter-level accuracy in sensing and localization~\cite{thz_isac_importance}. Utilizing the terahertz (THz) band, which spans from $100$\,GHz to $10$\,THz, is considered as a promising approach to reach this goal. The vast and untapped spectrum resources and the short wavelengths of the THz band offer the potential to achieve ultra-precise sensing and localization capabilities in 6G communications~\cite{dupleich2024characterization,wu2021thz}.

Accurate channel modeling for sensing is essential for the effective design and performance evaluation of future ISAC systems. In the literature, channel measurements for sensing can be divided into two types, i.e., monostatic sensing and bistatic sensing. In monostatic sensing, the transmitter (Tx) and receiver (Rx) are positioned at the same location during measurements, whereas in bistatic sensing, the Tx and Rx are placed at different locations~\cite{sensing_case}. Recent advancements have been made in sensing channel studies at lower frequency bands, specifically in sub-$6$\,GHz and millimeter-wave (mmWave) bands~\cite{low_freq_isac3,low_freq_isac1,isac_jhzhang,low_freq_isac2,low_freq_isac4}. In~\cite{low_freq_isac3}, measurement-based channel models for monostatic sensing at $2$-$8$\,GHz are presented. Monostatic sensing channel measurements are conducted in an outdoor scenario at $10$\,GHz and a geometry-based stochastic ISAC channel model is proposed in~\cite{low_freq_isac1}. Communication and monostatic sensing channel measurements at $28$\,GHz band are conducted in~\cite{isac_jhzhang} for an indoor hall scenario. A shared-cluster-based stochastic model is proposed for millimeter-wave (mmWave) ISAC channels. MmWave measurement campaigns in the monostatic sensing case are conducted in an indoor corridor scenario in~\cite{low_freq_isac2}. The environment mapping results are then presented, demonstrating the excellent
prospects of radio-based environment sensing and mapping
in future mm-wave networks. In~\cite{low_freq_isac4}, bistatic sensing channel measurements are performed in an indoor laboratory scenario at $24$\,GHz. The Doppler characteristics are then presented. The limited spectrum resources in lower frequency bands constrain the ranging accuracy for environment mapping and reconstruction.

As noted earlier, THz frequencies are expected to achieve much higher-ranging accuracy. However, only a few studies have explored THz sensing channels. In~\cite{hybrid_isac}, $140$\,GHz bistatic and monostatic sensing channel measurement campaigns are carried out in both indoor and outdoor scenarios. Moreover, a ray-tracing-stochastic hybrid model for the ISAC channel is proposed for $140$\,GHz. Bistatic sensing channel measurements were conducted in an industrial scenario at a frequency of $190$\,GHz, as reported in~\cite{dupleich2024characterization}. By leveraging high delay and angular resolution, the scatterers in the industrial environment can be effectively detected. In~\cite{thz_mono_sen_ybli}, channel measurements are carried out at $306$-$321$\,GHz bands in three different indoor scenarios, i.e., empty room, hallway, and corridor. Ranging accuracy and back-scattering coefficient are calculated and compared for these three scenarios. In~\cite{lotti2023radio}, monostatic sensing measurements are conducted in an indoor laboratory scenario at $235$-$320$\,GHz. A radio-simultaneous localization and mapping (R-SLAM) algorithm is implemented to map the environment and track the transceiver locations. 
 
%Description of this work
Despite substantial efforts in THz sensing measurements and modeling, several important aspects are still inadequately addressed in the current literature: 1) As the wavelength decreases from lower frequency bands to the THz band, the propagation mechanisms differ significantly from those observed in lower-frequency channels. However, previous studies have not investigated the propagation mechanisms specific to THz sensing channels. 2) In the monostatic sensing case, the complex radiation pattern of the transceiver (TRx), which is required for a high-resolution parameter estimation (HRPE) algorithm, is hard to obtain in the THz band. 3) Literature~\cite{thz_mono_sen_ybli,lotti2023radio} have found that in some special geometry of the indoor scenario, i.e., corner, the monostatic sensing fails to detect and reconstruct the geometry. 4) Several state-of-the-art studies~\cite{low_freq_isac1,hybrid_isac} have identified both target-related and environment-related components, proposing preliminary hybrid models for ISAC channels at $10$ and $130$\,GHz. However, comprehensive sensing channel models for indoor monostatic sensing at higher frequencies have yet to be fully explored.

To fill the aforementioned gaps, in this work, THz vector network analyzer (VNA)-based channel measurements are conducted in an indoor laboratory environment for monostatic sensing. The measurements cover a frequency range of $290$-$310$\,GHz. A total of $28$ measurement locations are deployed. The directional scanning sounding (DSS) technique is used to obtain the sensing-channel spatial profiles. The power-angle-delay profiles (PADP) are analyzed, revealing distinct structural features such as the flat wall, corner, and window. Moreover, a low-complexity HRPE method for antenna de-embedding is proposed. This method leverages the element-wise space-alternating generalized expectation-maximization (SAGE) algorithm, combined with geometry and the multipath component distance (MCD), to classify multipath components (MPCs) with identical radiation pattern effects and subsequently de-embed the influence of the antenna's radiation pattern. The de-embedded MPC parameters are subsequently utilized for monostatic-sensing-based channel characterization. Finally, a hybrid channel model is proposed for THz monostatic sensing. The key contributions of this paper can be listed as follows:
\begin{itemize}
\item {\bf Ultra-precise sensing measurements}: Our measurements achieve ultra-precise temporal and angular resolution, approximately $0.05$\,ns and $1^{\circ}$, respectively, thanks to the broad bandwidth and fine rotation step used in the setup.
\item {\bf Low-complexity HRPE algorithm for antenna de-embedding}: The proposed algorithm first utilizes an element-wise SAGE algorithm to extract MPC parameters, such as amplitude and delay. Following this, a geometry-based MPC trajectory tracking algorithm is applied to de-embed the effects of the antenna's radiation pattern, enabling the extraction of MPCs from the pure propagation channel.
\item {\bf Distinguishing target-related and environment-related components}: The propagation mechanisms for monostatic sensing are analyzed by categorizing target-related MPCs (e.g., specular reflections) and environment-related MPCs (e.g., diffuse reflections) based on different geometrical features such as flat wall and corner.
\item {\bf Hybrid sensing model}: In this hybrid channel model, the target-related component of the channel is leveraged for target and material identification, while the environment-related component is used for geometric environment mapping and reconstruction. To illustrate this, a demonstration of geometric environment reconstruction and an analysis of reflection losses based on the estimated MPCs are provided.
\end{itemize}

The remainder of this paper is structured as below. Section~\ref{sec:sounder} outlines the monostatic-sensing-based measurement setup and scenario. The data processing, analysis, and geometry-based channel estimation algorithm are described in Section~\ref{sec:estimator}. The hybrid channel model for the monostatic sensing case is proposed in Section~\ref{sec:modeling}. Finally, concluding remarks are presented in Section~\ref{sec:conclusion}.

\begin{figure}
    \centering
    \subfigure []{\includegraphics[width=1\columnwidth]{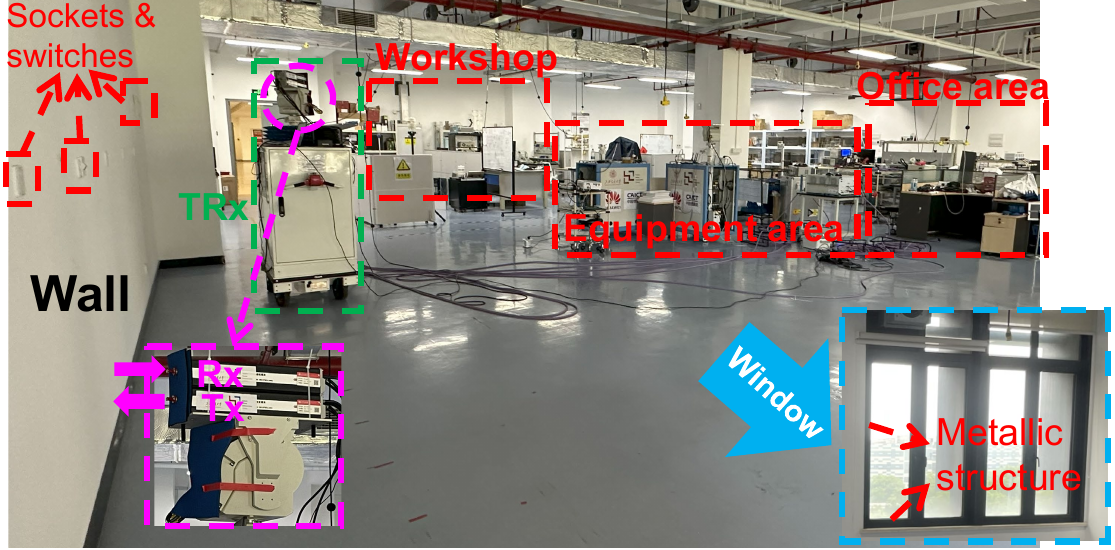}}
    \subfigure []{\includegraphics[width=1\columnwidth]{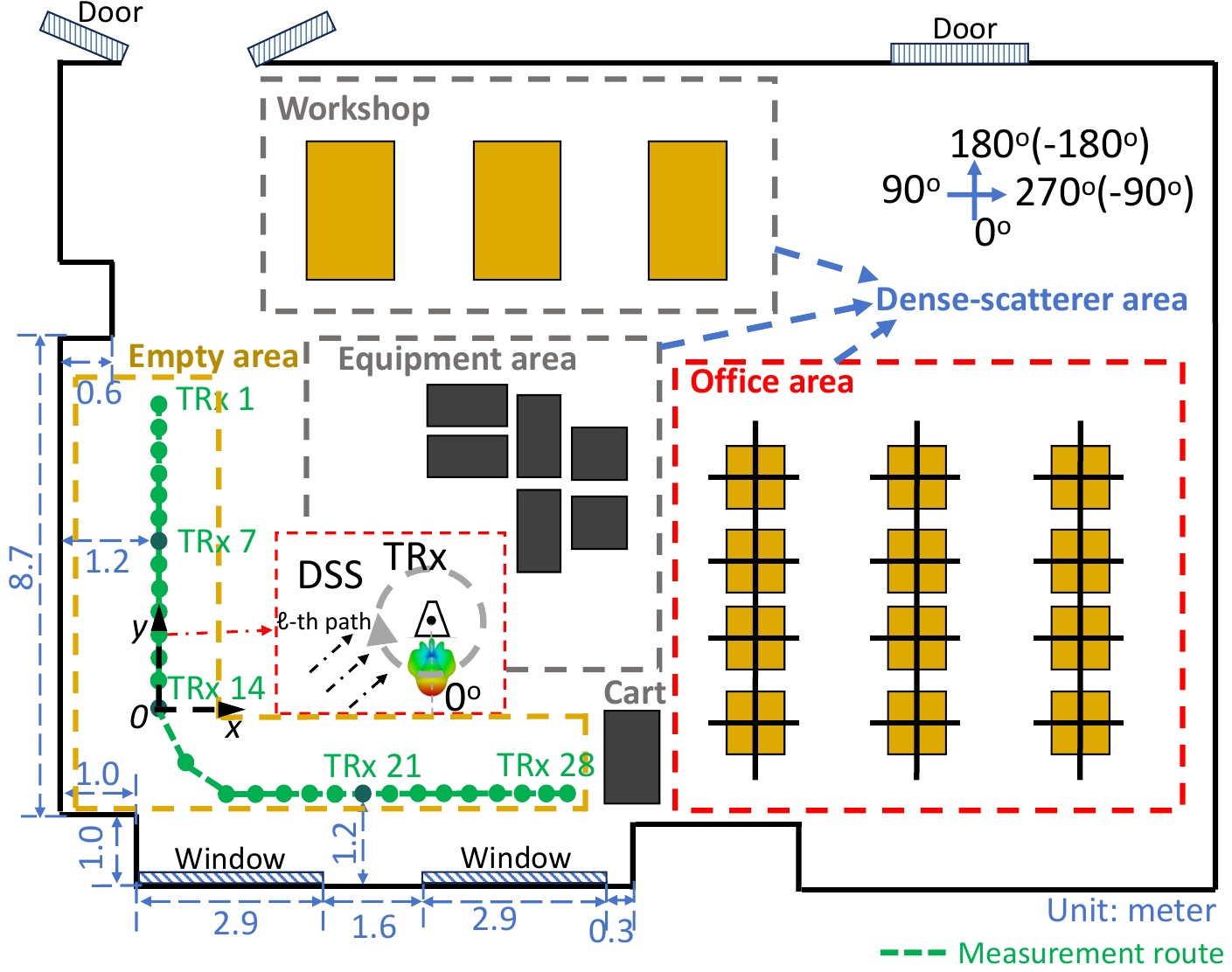}}
\caption {Diagram and photos of the measurement scenario for THz monostatic sensing. (a) Real photo. (b) Scenario diagram.}\label{fig:meas_pic}
\end{figure}

\section{Measurement Campaign}\label{sec:sounder}
\subsection{Sounder and Measurement Setup}
The measurement setup for monostatic sensing is depicted in Fig.~\ref{fig:meas_pic}~(a). In this study, a VNA-based channel sounder is used~\cite{lyu_vna,yuanbo_sage} covering a frequency range of $290$-$310$\,GHz. With a broad bandwidth of $20$\,GHz, an ultrahigh distance resolution of approximately $1.5$\,cm is achieved. The number of frequency points is set to $2001$, enabling the detection of distances up to $30$\,m. To ensure a high dynamic range in the measurements, the intermediate frequency (IF) bandwidth is set to $1$\,kHz. As illustrated in Fig.~\ref{fig:meas_pic}\,(a), the Tx and Rx are mounted together on a turntable to mimic monostatic sensing, similar to~\cite{thz_mono_sen_ybli}. The Tx and Rx employ the same horn antenna with $25.5$\,dBi antenna gain and $8^{\circ}$ half-power beamwidth (HPBW). Additionally, the directional scanning sounding (DSS) scheme is used to capture the spatial channel profiles in the azimuth plane. Note that the azimuthal radius (i.e., the distance between the phase center of the antenna to the rotation center) is $0.2$\,m. The TRx rotates within an azimuth angle range of $[0^{\circ}, 359^{\circ}]$, with a step size of $1^{\circ}$. The heights of the Tx and Rx are $2.0$\,m and $2.1$\,m, respectively. Prior to the measurements, back-to-back calibration is performed to remove system responses. Due to the extended sweep time of the VNA, each Tx-Rx (TRx) location measurement takes $1$ hour. The measurement configurations are summarized in Table~\ref{tab:configuration}.

\begin{table}
\centering
\caption{The measurement configuration for monostatic-sensing.}\label{tab:configuration}
\begin{tabular}{c|c}
\hline
Parameter   &  Value\\
\hline
\hline
Frequency range & $290$-$310$ GHz\\
\hline
Frequency bandwidth      & $20$ GHz\\
\hline
Space resolution     &  $15$\,mm\\
\hline
Frequency point      & $2001$\\
\hline
Maximum detectable distance &  $30$\,m\\
\hline
IF bandwidth            & $1$ kHz\\
\hline
Transmitted power & $10$ dBm\\
\hline
Tx \& Rx antenna type   & Horn\\
\hline
Antenna gain   & $25.5$ dBi\\
\hline
Azimuthal HPBW           & $8^{\circ}$\\
\hline
Rotation step     & $1^{\circ}$\\
\hline
Azimuthal rotation range  & $[0^{\circ}:359^{\circ}]$\\
\hline
Tx Antenna height     & $2.0$ m\\
\hline
Rx Antenna height     & $2.1$ m\\
\hline
\end{tabular}
\end{table}

% \begin{table}
% \centering
% \caption{The measurement configuration for monostatic-sensing.}\label{tab:configuration}
% \begin{tabular}{c|c}
% \hline
% Parameter   &  Value\\
% \hline
% \hline
% Frequency & $140$ GHz\\
% \hline
% Frequency bandwidth      & $1.536$ GHz\\
% \hline
% Delay resolution     &  $0.65$\,ns\\
% \hline
% Space resolution     &  $0.2$\,m\\
% \hline
% Measuring speed &  $< 2$ ms per CIR\\
% \hline
% Tx \& Rx antenna type   & Horn\\
% \hline
% Antenna gain   & $25$-$27$ dBi\\
% \hline
% Azimuthal HPBW           & $8^{\circ}$\\
% \hline
% Rotation step     & $5^{\circ}$\\
% \hline
% Azimuthal rotation range  & $[0^{\circ}:355^{\circ}]$\\
% \hline
% Elevational rotation range  & $[-20^{\circ}:20^{\circ}]$\\
% \hline
% Tx Antenna height     & $1.8$ m\\
% \hline
% Rx Antenna height     & $1.8$ m\\
% \hline
% \end{tabular}
% \end{table}

\subsection{Measurement Scenario}
The indoor scenario and the deployment of the TRx are illustrated in Fig.~\ref{fig:meas_pic}~(b). The channel measurements are performed in an indoor laboratory scenario. The scenario can be divided into four areas, i.e., workshop, equipment area, office area, and empty area. Note that the workshop, equipment area, and office area can be regarded as dense-scatterer areas in the measurements since there are many metallic, plastic, and wooden objects in these areas. In order to reduce interference and measurement uncertainty, our measurement campaign is conducted in the empty area with an ``L''-shape route. During the measurements, $28$ TRx locations are measured, with $0.5$\,m distance separation between the adjacent TRx locations. The distance between the TRx location and the wall structure are set to be the same as $1.2$\,m. The left-side wall features several sockets and switches, while the bottom wall contains two metallic-framed windows. 

\section{Data Processing and Analysis}\label{sec:estimator}
\begin{figure}
    \centering
    \subfigure []{\includegraphics[width=1\columnwidth]{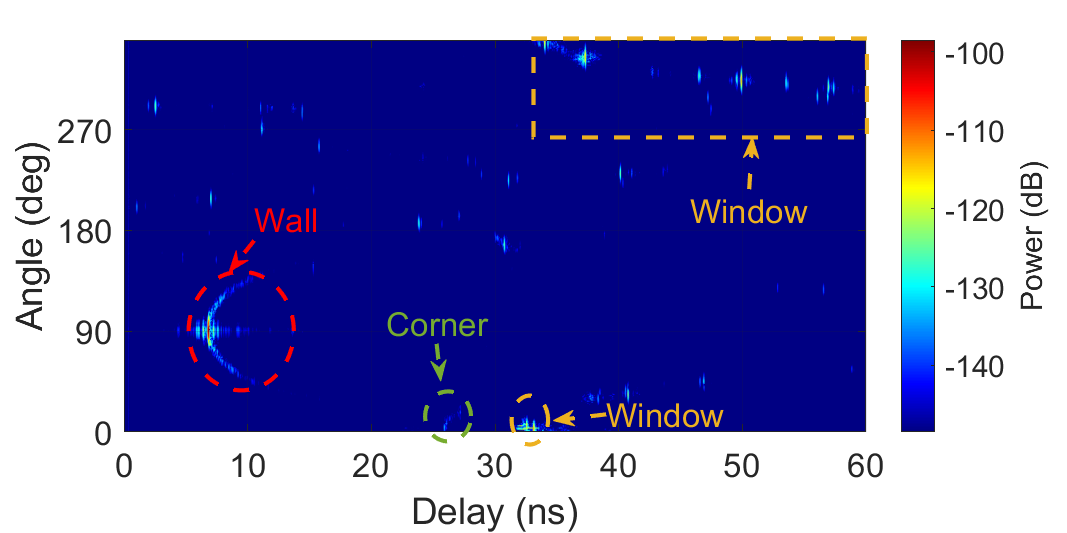}}
    \subfigure []{\includegraphics[width=1\columnwidth]{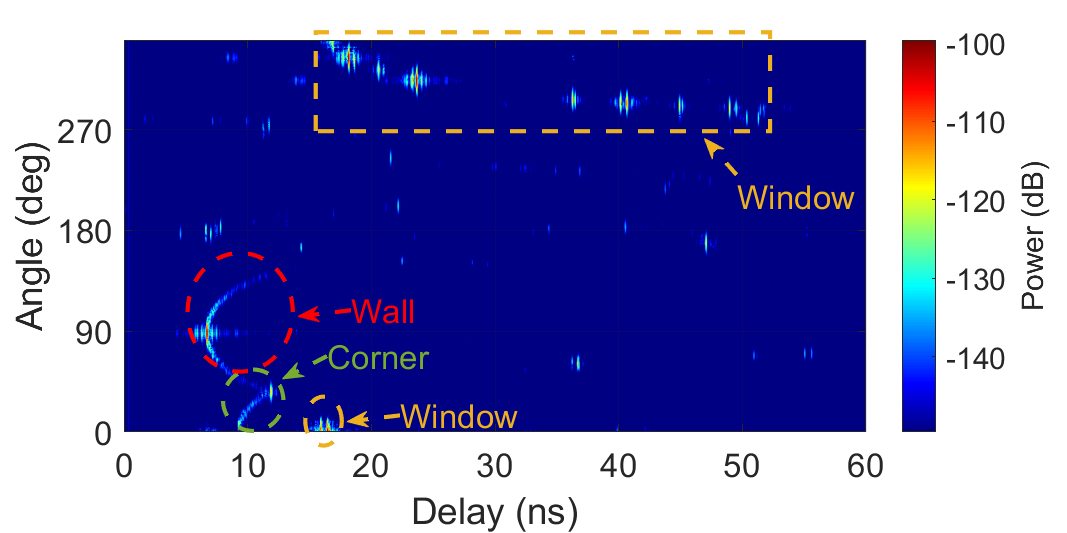}}
    \subfigure []{\includegraphics[width=1\columnwidth]{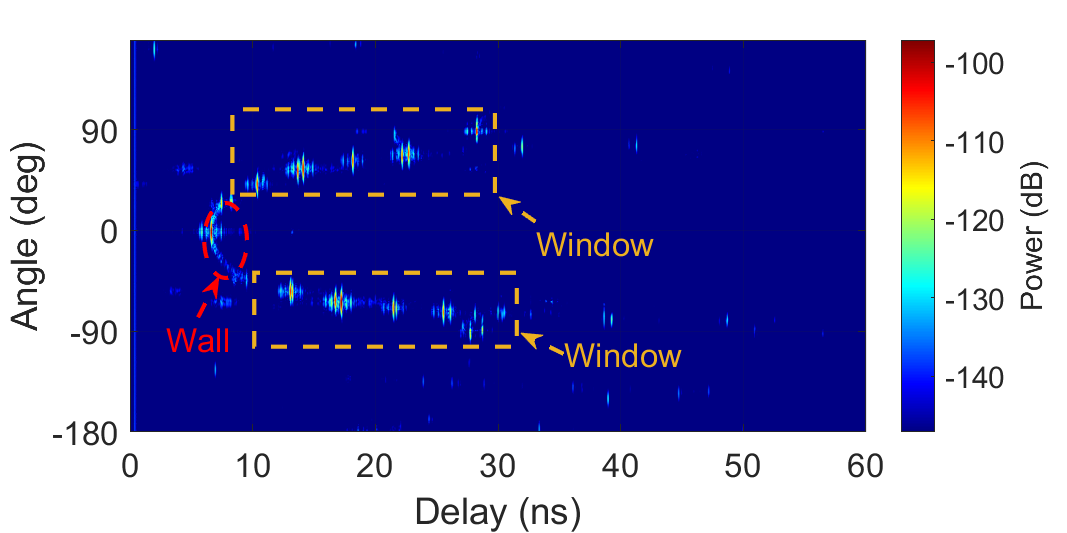}}
\caption {Exemplary PADPs from the measurements. (a) TRx\,$7$. (b) TRx\,$14$. (c) TRx\,$21$.}\label{fig:raw_results}
\end{figure}

In the measurements, directional channel frequency responses (CFRs) $H_{m}(f,\varphi)$ were obtained for each monostatic-sensing location, where $m$, $f$, and $\varphi$ represent the index of the TRx location, the carrier frequency, and the azimuthal rotation angle. The power-angle-delay profiles (PADPs) $p_{m}(\tau,\varphi)$ can be obtained through the application of the inverse discrete Fourier transform (IDFT), where $\tau$ indicates the propagation delay. The exemplary PADPs results at TRx\,$7$, TRx\,$14$, and TRx\,$21$ are depicted in Fig.~\ref{fig:raw_results}. By employing $20$\,GHz bandwidth, corresponding to approximately $0.05$\,ns delay resolution and $15$\,mm space resolution, the MPCs can be distinguished in the delay domain. In Fig.~\ref{fig:raw_results}~(a), strong reflection from the flat wall in the azimuth range of $[41^{\circ},140^{\circ}]$ is well-detected. It can be also seen that the TRx receives strong reflections from the metallic structure of the window in the azimuth range of $[0^{\circ},13^{\circ}]$ and $[288^{\circ},359^{\circ}]$. Moreover, weak reflection from the corner can be detected at TRx\,$7$. As for the PADP for TRx\,$14$ in Fig.~\ref{fig:raw_results}~(b), the reflection from the flat wall, corner, and window can be clearly distinguished. In the PADP for TRx\,$21$ in Fig.~\ref{fig:raw_results}~(c), since the TRx\,$21$ is located in the middle of the two windows, the reflections from the two windows and the bottom wall in Fig.~\ref{fig:meas_pic}~(b) can be well-detected. Furthermore, many reflections from the dense scatterer area (i.e., workshop, equipment area and office area) are observed in the azimuth range of $[160^{\circ},300^{\circ}]$, $[148^{\circ},280^{\circ}]$, and $[154^{\circ},245^{\circ}]$ at the three TRx locations in Fig.~\ref{fig:raw_results}, respectively. We also calculate the geometric locations and relationships of the wall, corner, and window structures based on the propagation delays and azimuth angles, and compare them with the real-world scenario, demonstrating a strong match between them.

\subsection{Signal model}
The directional CFR in the rotation angle of $\varphi$ in the $m^{\rm th}$ location can be expressed as
\begin{align}
H_{m}(f,\varphi) = \sum_{\ell=1}^{L} \alpha_{\ell} \exp(-j2\pi f \tau_{\ell}) \cdot a_{\rm TRx}(f,\varphi),\label{equ:signal}
\end{align}
where $\alpha_{\ell}$, $\tau_{\ell}$, and $f$ denote the complex amplitude, delay of the $\ell^{\rm th}$ path, and the carrier frequency, respectively. $L$ represents the total number of the MPCs. $a_{\rm TRx,\ell}(f,\varphi)$ is the rotation manifold coefficient, which contains the phase difference caused by DSS and the antenna pattern for $\ell^{\rm th}$ path. The coefficient $a_{\rm TRx}(f,\varphi)$ can be written as
\begin{align}
a_{\rm TRx,\ell}(f,\varphi) = \left[\exp(j2\pi f r \cos(\varphi)/c)\right]^{2} \cdot G_{\rm TRx}(f,\phi_{\ell}-\varphi),
\end{align}
where $c$ and $r$ represent the speed of light and the azimuth radius, respectively. In this work, we assume that the departure azimuth angle and impinge azimuth angle are the same as $\phi_{\ell}$ for the $\ell^{\rm th}$ path. $G_{\rm TRx}(f,\phi_{\ell}-\varphi)$ is the complex radiation pattern of the mimic monostatic sensor. As discussed in literature~\cite{yuanbo_sage,lyu_phase_wband}, the complex radiation pattern is hard to obtain due to the system non-idealities and uncertainties, e.g. antenna phase center positioning error and cable movement. Besides, the different heights of the Tx and Rx introduce additional phase differences to the radiation pattern $G_{\rm TRx}(f,\phi_{\ell}-\varphi)$. Thus, according to the aforementioned discussion, the HRPE algorithms and antenna de-embedding methods are not suitable for THz channel parameter estimation in the monostatic sensing case.

\subsection{Channel Parameter Estimation and Analysis}\label{sec:hybird_def}
The HPBW of the antennas is narrow (i.e., $8^{\circ}$) and the scanning step is $1^{\circ}$. These configurations result in a high angular resolution in our measurement results, which can be illustrated from the PADP results in Fig.~\ref{fig:raw_results}. Thus, to reduce the algorithm complexity, in this work, an element-wise SAGE algorithm is employed~\cite{bernald_sage}. During the SAGE process, a signal threshold set at $10$\,dB higher than the noise floor \cite{kguan_intrawagon} is applied to estimate the multipath components (MPCs). The frequency response $H_{m,\varphi}(f)$ at the certain rotation angle $\varphi$ in the $m^{\rm th}$ TRx location can be simplified as
\begin{align}
H_{m,\varphi}(f) = &\sum_{\ell=1}^{L_{\varphi}} \alpha^{(m,\varphi)}_{\ell} \exp(-j2\pi f \tau^{(m,\varphi)}_{\ell}),\label{equ:signal2}
\end{align}
where $L_{\varphi}$, $\alpha^{(m,\varphi)}_{\ell}$, and $\tau^{(m,\varphi)}_{\ell}$ denote the total number of MPCs, amplitude and delay of $\ell^{\rm th}$ path at the certain rotation angle $\varphi$ in the $m^{\rm th}$ TRx location, respectively.

In the channel parameter estimation, we need to determine a set of parameters for $m^{\rm th}$ TRx location denoted as $\varTheta^{(m)} = [\varTheta^{(m)}_{1}, \varTheta^{(m)}_{2}, \cdots, \varTheta^{(m)}_{359}]$. Each $\varTheta^{(m)}_{\varphi}$ is a matrix containing the parameters for the MPCs in the rotation angle of $\varphi$ with a total number of $L_{\varphi}$, which can be expressed as 
 \begin{align}
\varTheta^{(m)}_{\varphi}=\left[ \begin{matrix}
\alpha _{1}^{(m,\varphi)}&	\cdots &    \alpha _{L\varphi}^{(m,\varphi)}\\
\tau _{1}^{(m,\varphi)}&    \cdots &    \tau _{L\varphi}^{(m,\varphi)}\\
\end{matrix} \right],
 \end{align}
 
\begin{figure}
\centering
\includegraphics[width=0.8\columnwidth]{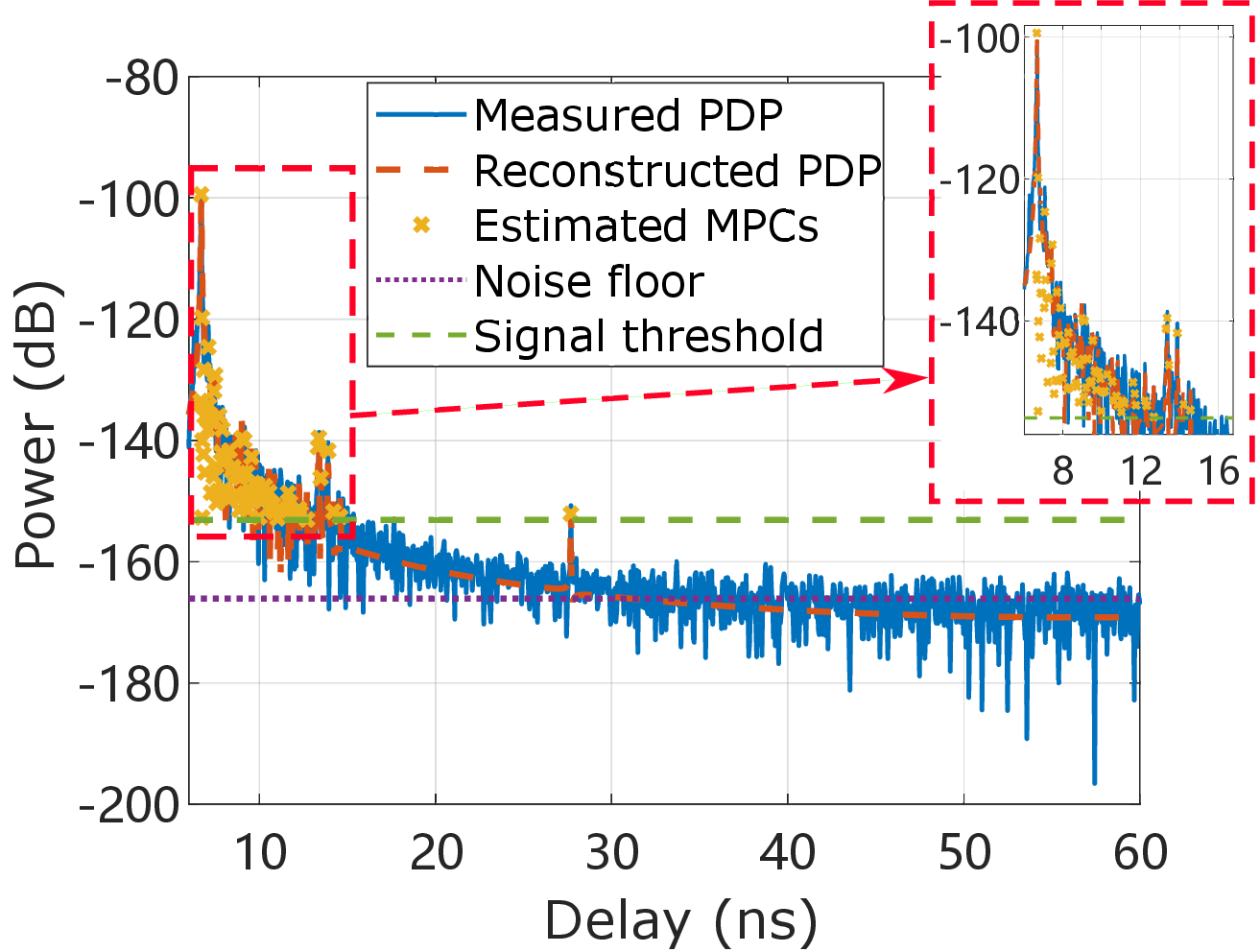}
\caption {Exemplary reconstructed PDP from estimated results compared with the measurements.}\label{fig:sage_results}
\end{figure}

The exemplary result in the power-delay profile (PDP) is shown in Fig.~\ref{fig:sage_results}. It can be observed that all the MPCs above the signal threshold are extracted from the measurement data. We also compare the reconstructed PDP based on (\ref{equ:signal2}) and the comparison shows a good agreement between the measured and reconstructed PDP.

\begin{figure}
    \centering
    \subfigure []{\includegraphics[width=0.99\columnwidth]{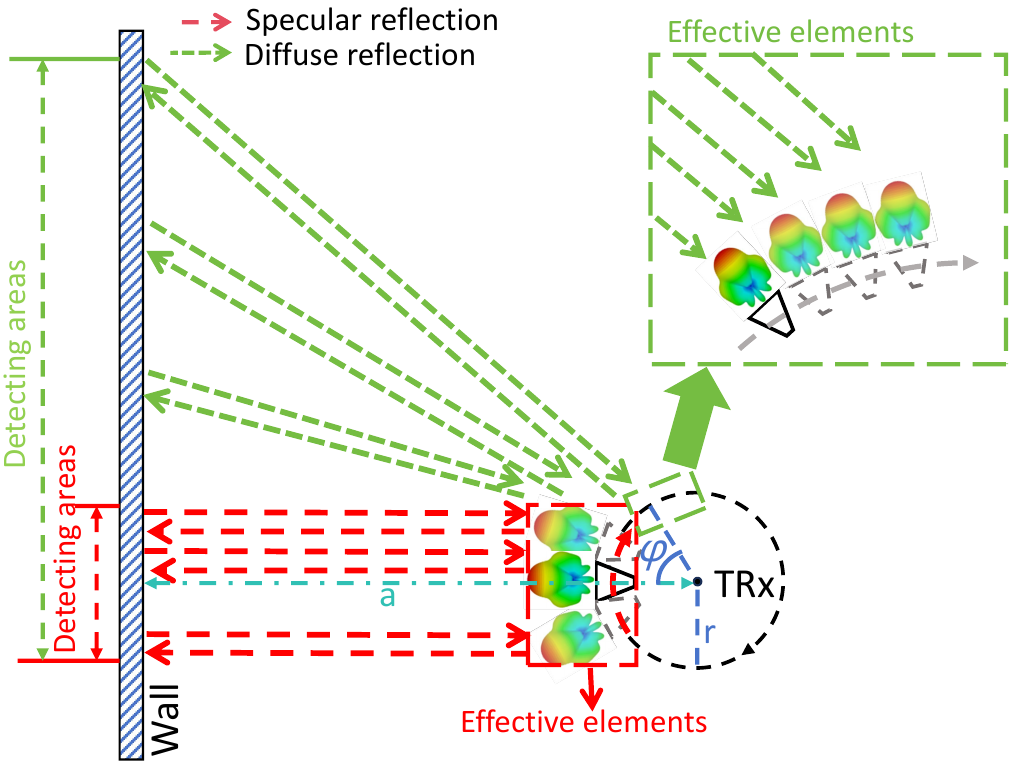}}
    \subfigure []{\includegraphics[width=0.8\columnwidth]{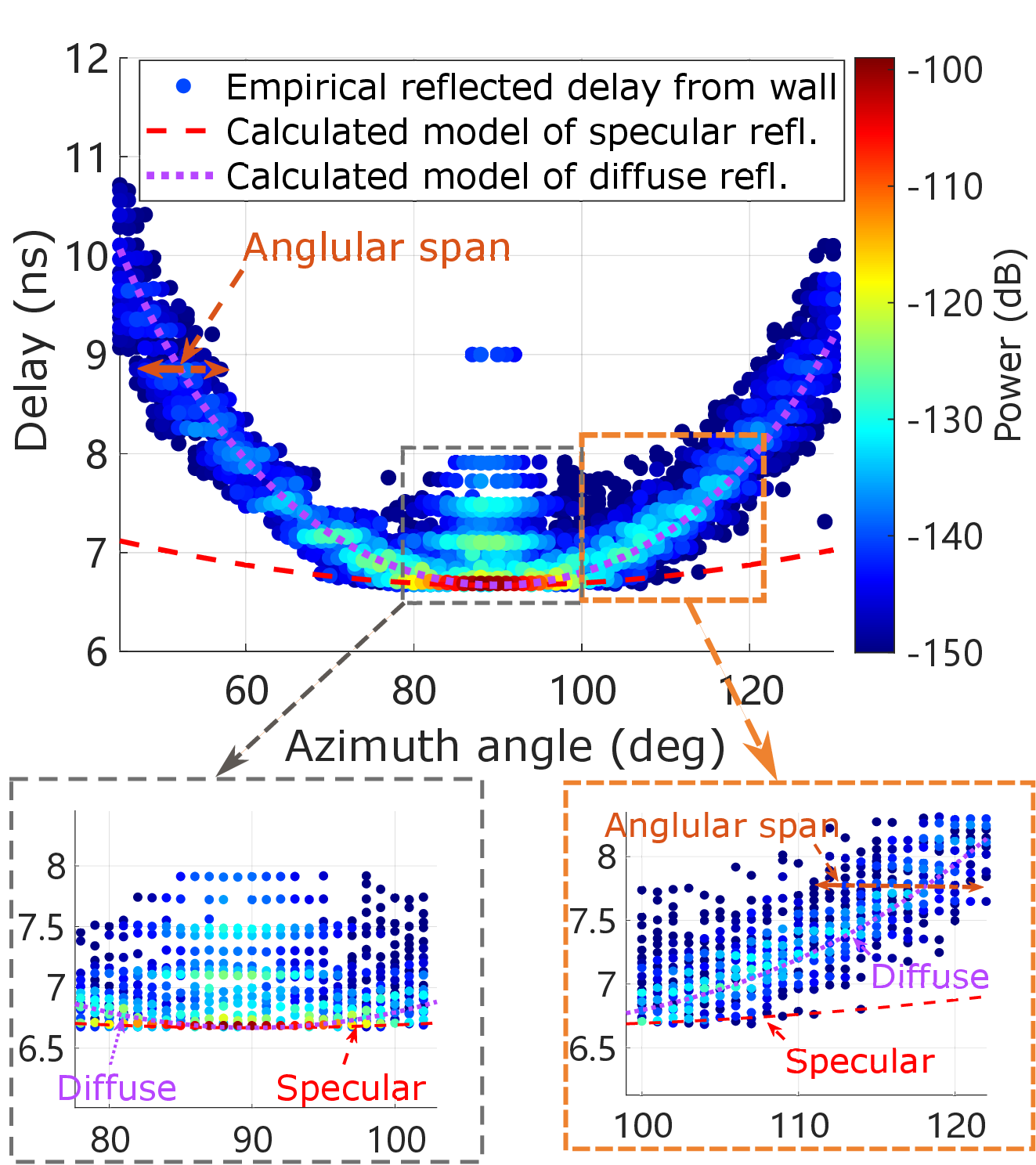}}
\caption {Comparison of MPC for flat wall reflections with the geometric model. (a) Comparison of geometric specular and diffuse reflections with the scenario. (b) Relationship of the extracted MPC with the geometric model.}\label{fig:padp_compare_wall}
\end{figure}

% Specular and diffuse reflection identification
To develop a hybrid channel model, it is crucial to differentiate between the target-related components (i.e., specular reflections) and the environment-related components (i.e., diffuse reflections). In our work, both specular reflections and diffuse reflections are distinctly identified in the estimated results. Taking the flat wall as an example on the azimuth angle range of $[40^{\circ},130^{\circ}]$ in the TRx\,$14$, we first calculate and fit the geometrical models on the delay domain based on specular reflection and diffuse reflection assumption, as depicted in Fig.~\ref{fig:padp_compare_wall}~(a). The propagation delay of the specular reflection from the flat wall $\tau_{\rm w,spec}$ can be calculated as
\begin{align}
\tau_{\rm w,spec}= \frac{2\left[a-r\left(1-\cos(\varphi)\right)\right]}{c},
\end{align}
where $a$ denotes the distance from the rotation center to the flat wall.

While the propagation delay of the diffuse reflection $\tau_{\rm diff}$ is expressed as
\begin{align}
\begin{split}
\tau_{\rm diff} = \frac{2[a/\cos(\varphi)-r]}{c},\label{eq:diff_model}
\end{split}
\end{align}

In Fig.~\ref{fig:padp_compare_wall}~(b), the delay variation with azimuth changes for the specular reflection model is significantly smaller than that for the diffuse reflection model. Besides, as illustrated in Fig.~\ref{fig:padp_compare_wall}~(a), the diffuse reflection can be used for detecting and sensing larger areas compared to the specular reflection. The delay model of the specular reflection is observed to vary in a delay range from $6.67$ to $7.12$\,ns in this angle range of $[40^{\circ},130^{\circ}]$, while the delay model of the diffuse reflection spans a larger delay range of $[6.71,10.66]$\,ns. Compared with the estimated MPC results in Fig.~\ref{fig:padp_compare_wall}~(b), it can be observed that the MPCs in the azimuth range of $[69^{\circ},114^{\circ}]$ and the delay range of $[6.68, 6.80]$\,ns fit well with the specular delay model, while most of the estimated MPCs in this azimuth range of $[40^{\circ},130^{\circ}]$ align well with the geometry-based diffuse model, which demonstrates the feasibility of establishing a hybrid model for monostatic sensing case. In this approach, specular reflections are modeled as the target-related part of the sensing channel, while diffuse reflections are treated as the environment-related component. Note that the angle span of the extracted diffuse MPCs across the geometry-based model is observed, which is mainly caused by the radiation pattern of the employed antennas.

\begin{figure}
    \centering
    \subfigure []{\includegraphics[width=0.7\columnwidth]{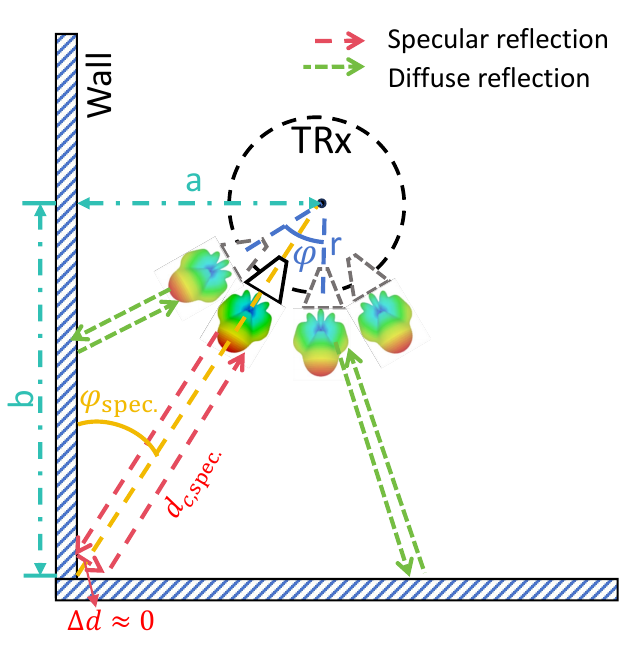}}
    \subfigure []{\includegraphics[width=0.8\columnwidth]{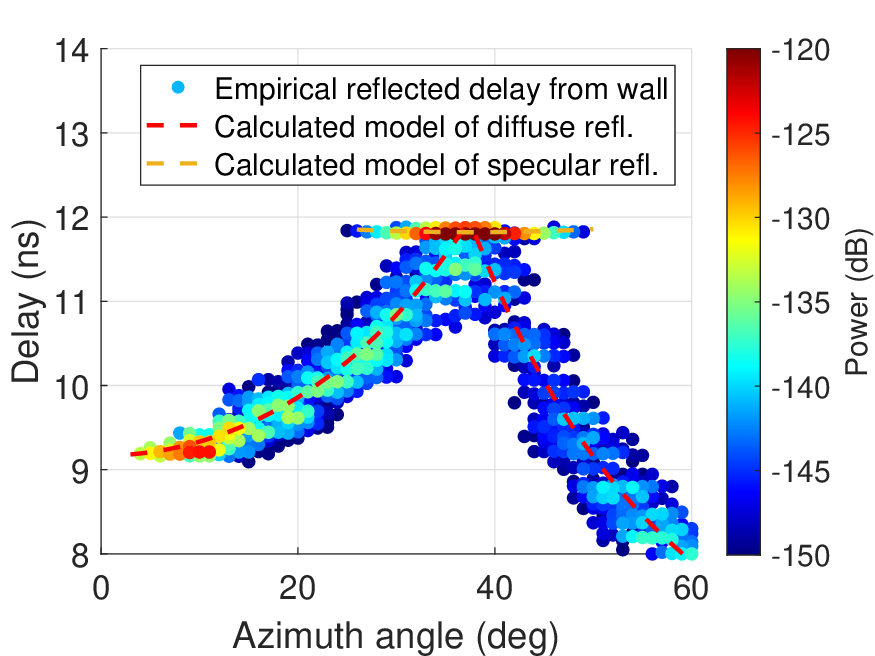}}
\caption {Comparison of MPCs for corner reflections with the geometric model. (a) Comparison of geometric specular and diffuse reflections with the scenario. (b) Relationship of the extracted MPC with the geometric model.}\label{fig:padp_compare_corner}
\end{figure}

As noted in the introduction, detecting corners, which are common structures in indoor environments, poses a challenge in monostatic sensing cases. Here, a geometry-based method is applied to model the corner. The geometric model of the reflection from the corner and the exemplary extracted MPC results in the azimuth angle range of $[0^{\circ},60^{\circ}]$ at TRx\,$14$ compared with the geometry are illustrated in Fig.~\ref{fig:padp_compare_corner}. Fig.~\ref{fig:padp_compare_corner}~(a) demonstrates the reflection mechanism of monostatic sensing from the corner. Similar to the propagation MPCs in the wall case, the reflection propagation from the corner consists of a specularly reflected path and multiple diffuse reflected paths. Based on the geometry, the specular one is a second-order reflection, and the propagation delay of it $\tau_{\rm c,spec}$ can be calculated as:
\begin{align}
\begin{split}
\tau_{\rm c,spec} &= 2 \cdot d_{\rm c,spec} + \Delta d \approx 2 \cdot d_{\rm c,spec}  \\
&= \frac{2[a/\sin(\varphi_{\rm spec})-r\cos(\varphi)]}{c},
\end{split}
\end{align}
where $d_{\rm c,spec}$ and $\Delta d$ denote the distance from the antenna element to the corner and the distance from the first-order reflection point to the second-order reflection point, respectively. $\varphi_{\rm spec}$ indicates the angle between the wall and the line extending from the TRx rotation center to the corner. The calculation of the diffuse reflected propagation delay $\tau_{\rm diff}$ is the same as (\ref{eq:diff_model}) in the flat wall case. Fig.~\ref{fig:padp_compare_corner}~(b) shows the extracted MPC results compared with the geometric models. Due to the azimuthal antenna pattern effect of the TRx, the specular reflection varies within an azimuth range of $[26^{\circ},50^{\circ}]$. Besides, due to the small azimuth radius, i.e., $0.2$\,m, the specularly reflected paths are observed to vary within a narrow delay range of $[11.79,11.82]$\,ns. For the diffuse reflection, the delay is seen to increase in the azimuth range of $[0^{\circ},38^{\circ}]$ and decrease in the azimuth range of $[38^{\circ},60^{\circ}]$. We also calculate the geometric-based delay models for both specular and diffuse reflections at the corner and compare them with the extracted MPC results, which demonstrate a strong match.

\subsection{Geometry-Based MPC Trajectory Tracking}
In this work, MPC trajectories in rotation angle can be used to classify the antenna pattern effect. The proposed MPC trajectory tracking algorithm is based on the geometry and multipath component distance (MCD), which is similar to~\cite{lyu_tracking}. The key difference between our proposed algorithm and previous work is that, in most communication scenarios, the line-of-sight (LoS) trajectory can be obtained and serves as a good reference for selecting the MCD threshold. However, in our case of monostatic sensing, no LoS trajectory is available as a reference, necessitating the use of an alternative approach, which will be explained later.

% Geometry
Based on the geometric delay model and the PADP results in Fig.~\ref{fig:padp_compare_wall} and Fig.~\ref{fig:padp_compare_corner}, the high path loss at THz frequencies and the narrow HPBW of the antennas cause the `birth' and `death' of MPCs, influenced by the antenna pattern, to occur within a narrow azimuth range. For instance, in Fig.~\ref{fig:padp_compare_corner}~(b), the specular reflected MPCs are observed in the azimuth range of $[26^{\circ},50^{\circ}]$. The diffuse reflected MPCs, which have weaker power, are impacted by the antenna pattern over an even narrower azimuth range. Furthermore, due to the small rotation radius and rotation step, i.e., $0.2$\,m and $1^{\circ}$, respectively, the delay change in the adjacent rotation angle can be neglected. Thus, the delay difference between the MPCs with the same antenna pattern in adjacent azimuth is insignificant. Based on these observations, a geometry-based trajectory tracking algorithm is proposed to track and categorize the MPCs caused by the radiation pattern effect. 

We first select the MPCs in the adjacent rotation angles with the delay difference between them within $\pm 0.02$\,ns. Then, to avoid choosing the wrong MPCs between the adjacent rotation angle, the MCD is employed to evaluate the separation between the MPCs. For $m^{\rm th}$ TRx location, the MCD between $\ell^{\rm th}$ estimated MPC at $\varphi$ rotation angle and $j^{\rm th}$ estimated MPC at $\varphi+\Delta\varphi$ rotation angle can be expressed as
\begin{align}
\mathrm{MCD}^{(m)}_{\varphi,\ell ,j}=\sqrt{\omega _p\left( \Delta p^{(m)}_{\varphi,\ell ,j} \right) ^2+\omega _{\tau}\left( \Delta \tau^{(m)}_{\varphi,\ell ,j} \right) ^2},
\end{align}
where $\omega_{p}$ and $\omega_{\tau}$ are the weight factor of the power and delay, respectively, which will be explained later. We define $\Delta p^{(m)}_{\varphi,\ell,j}$ and $\Delta\tau^{(m)}_{\varphi,\ell,j}$ as:
\begin{align}
\Delta p^{(m)}_{\varphi,\ell,j} = |p^{(m)}_{\varphi,\ell}-p^{(m)}_{\varphi+\Delta\varphi,j}|,
\end{align}
\begin{align}
\Delta \tau^{(m)}_{\varphi,\ell,j} = |\tau^{(m)}_{\varphi,\ell}-\tau^{(m)}_{\varphi+\Delta\varphi,j}|,
\end{align}
\begin{align}
p^{(m)}_{\varphi,\ell} = 10\cdot\log_{10}|\alpha^{(m)}_{\varphi,\ell}|^2,
\end{align}
 where $\hat{\alpha}^{(m)}_{\varphi,\ell}$ and $\hat{\tau}^{(m)}_{\varphi,\ell}$ are the estimated amplitude and delay of the $\ell^{\rm th}$ MPC at the rotation angle of $\varphi$ in the $m^{\rm th}$ TRx location, respectively.

\begin{figure}
    \centering
    \subfigure []{\includegraphics[width=0.8\columnwidth]{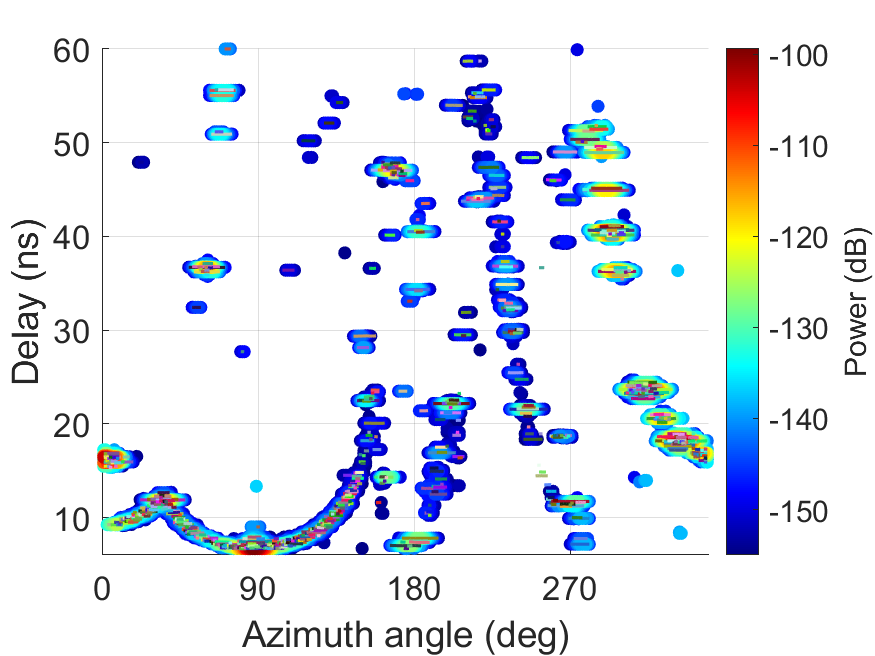}}
    \subfigure []{\includegraphics[width=1\columnwidth]{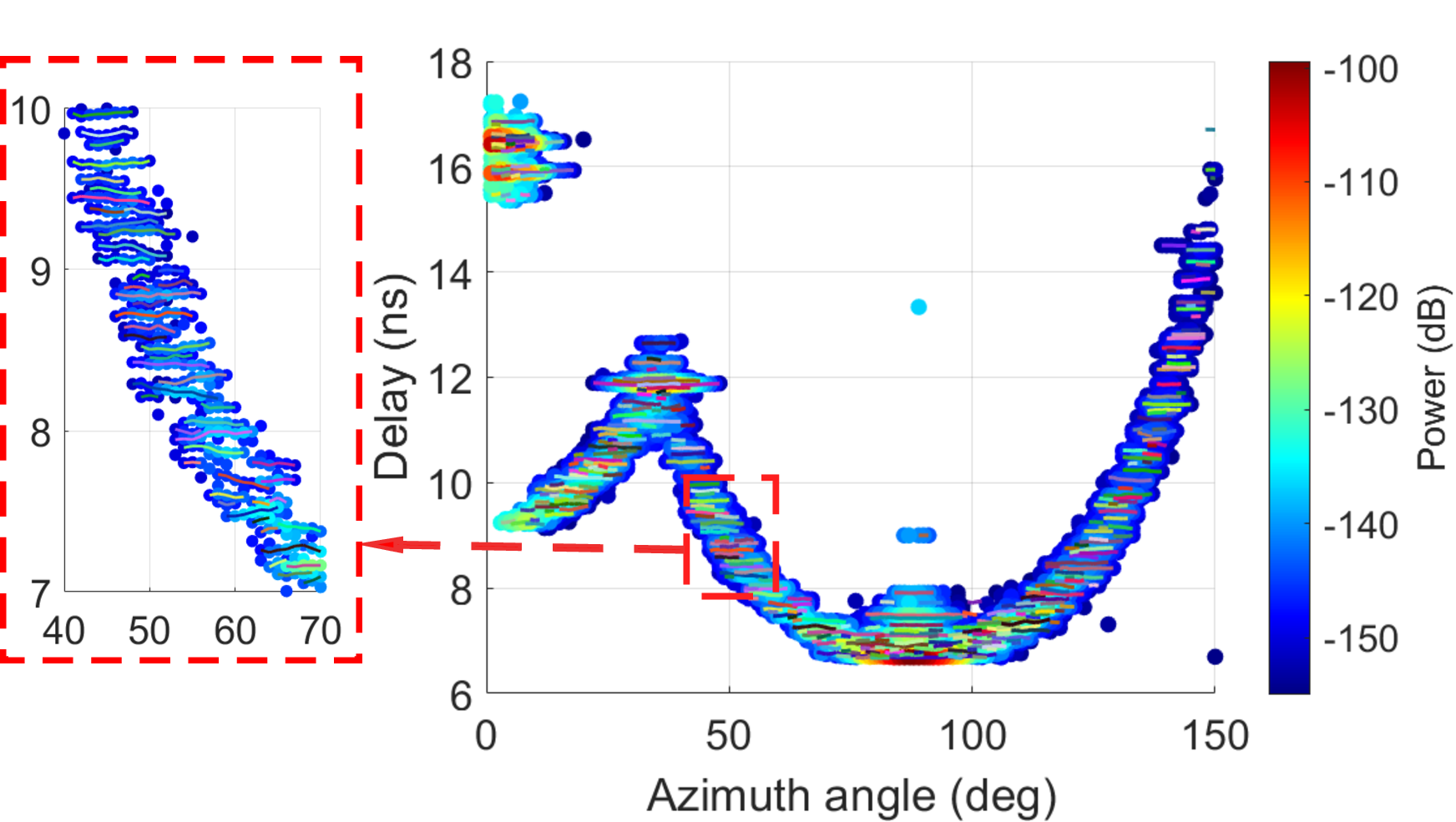}}
\caption {The tracked MPC trajectories compared with the extracted MPCs at TRx\,$14$ as an example. (a) Results in in the azimuth range of $[0^{\circ},360^{\circ}]$. (b) Zoomed-in results in the azimuth range of $[0^{\circ},150^{\circ}]$ .}\label{fig:traj_results}
\end{figure}

In our measurements for monostatic sensing, which is a non-line-of-sight (NLoS) case, LoS trajectories cannot be obtained for reference. Thus, in our case, the specular reflection trajectory from the wall can be used as an alternative reference for choosing appropriate $\omega_{p}$ and $\omega_{\tau}$. We express the specular reflection path at $m^{\rm th}$ Rx location as $\varTheta^{(m)}_{\mathrm{spec}}$. Assuming $x$ is one of the estimated parameter $\{p,\tau\}$, the weight $\omega_{x}$ can be calculated as
\begin{align}
\omega_{x} = \frac{1}{\sqrt{S_{x}}},
\end{align}
where $S_{x}$ is the unbiased sample standard deviation of the parameter $x$ as
\begin{align}
S_{x} = \sqrt{\frac{1}{N-2}\sum_{n=1}^{N-1}\left(\Delta x^{\rm spec}_{n}- \frac{\sum_{n=1}^{N-1}\Delta x^{\rm spec}_{n}}{N-1}\right)},
\end{align}
where $N$ is the total number of the rotation angle set.

The estimated MPC trajectories using the proposed algorithm are shown in Fig.~\ref{fig:traj_results} and compared with the extracted MPC results (taking TRx\,$14$ as an example). The total number of the extracted MPC trajectories at TRx\,$14$ is $774$. To show the clear results, we zoom in Fig.~\ref{fig:traj_results}~(a) in the azimuth range of $[0^{\circ},150^{\circ}]$, where the flat wall, corner, and window are clearly detected. It is evident that the trajectories of MPCs influenced by the radiation pattern are accurately classified.

\begin{figure}
\centering
\includegraphics[width=0.8\columnwidth]{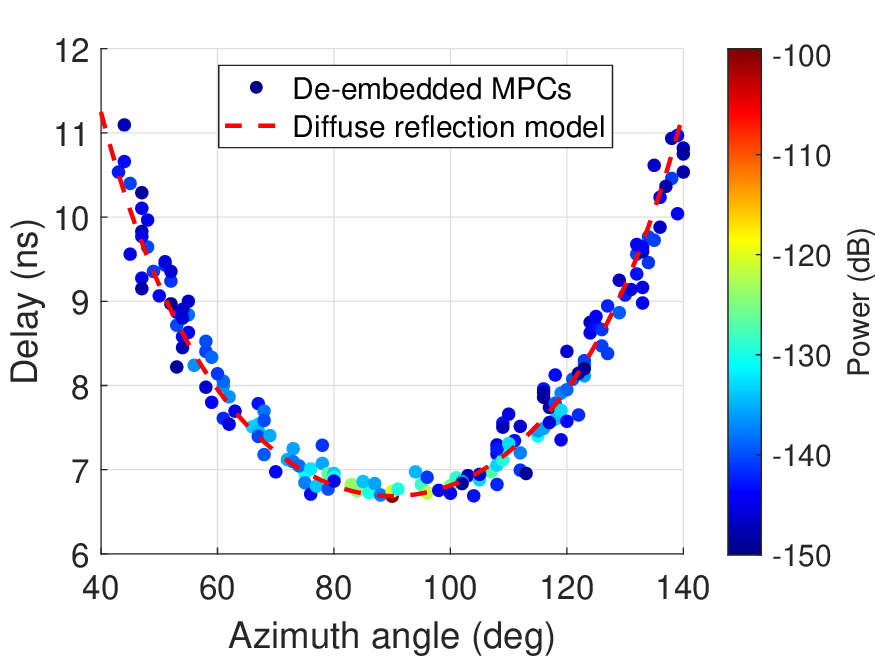}
\caption {The de-embedded MPCs compared with the geometry (in the azimuth range of $[40^{\circ},140^{\circ}]$ at TRx\,$14$ as an example).}\label{fig:deembed_mpc_results}
\end{figure}

% Assuming that the number of trajectories is $K_{m}$ in the $m$-th TRx location and the $k$-th extracted trajectory caused by the radiation pattern effect is $T_{k}$, the $k$-th trajectory in the azimuth range of $[\varphi_L,\varphi_H]$ can be expressed as
% \begin{align}
% T^{(m)}_{k}= ,
% \end{align}
% where $\varphi$

We can simply take the MPCs with the maximum power as the de-embedded MPCs in the trajectories. Assuming the total de-embeded MPC number is $L_{m}$ for the $m^{\rm th}$ TRx location, the de-embedded parameter set $\hat{\varTheta}^{(m)}_{\rm de}$ can be expressed as
 \begin{align}
\hat{\varTheta}^{(m)}_{\rm de}=\left[ \begin{matrix}
\hat{\alpha}_{1}^{(m)}&	\cdots & \hat{\alpha}_{\ell}^{(m)} &\cdots &  \hat{\alpha}_{Lm}^{(m)}\\
\hat{\tau}_{1}^{(m)}&    \cdots & \hat{\tau}_{\ell}^{(m)} &\cdots &    \hat{\tau}_{Lm}^{(m)}\\
\hat{\phi}_{1}^{(m)}&    \cdots & \hat{\phi}_{\ell}^{(m)} &\cdots &    \hat{\phi}_{Lm}^{(m)}\\
\end{matrix} \right],
 \end{align}
where $\alpha _{\ell}^{(m)}$, $\tau _{\ell}^{(m)}$, and $\phi_{\ell}^{(m)}$ denote the de-embedded amplitude, delay, and azimuth angle of $\ell^{\rm th}$ MPC, respectively.

Fig.~\ref{fig:deembed_mpc_results}~(a) illustrates the exemplary de-embedded MPCs compared with the geometric diffuse model, taking the azimuth range of $[40^{\circ},140^{\circ}]$ at TRx\,$14$ as an example. It is observed that the de-embedded MPCs match well with the geometry-based diffuse model using (\ref{eq:diff_model}). 
%The difference $\Delta \tau$ of the extracted MPC delay and the geometry-based model is also calculated and presented in the cumulative probability function (CDF), as depicted in Fig.\,\ref{fig:deembed_mpc_results}\,(b). The difference $\Delta \tau$ varies from $[-0.82,0.80]$\,ns and the CDF of $\Delta \tau$ fits well with the normal distribution of $\mathcal{N}(0,0.08)$. The observed delay variation is primarily attributed to the HPBW in elevation of the antennas and the roughness of the wall.

\section{Channel Characterization and Modeling}\label{sec:modeling}
In this section, a hybrid model for the monostatic sensing channel is proposed, where the target-related components contribute to target/material identification and the environment-related components can be used for geometrical environment reconstruction. 

\subsection{Hybrid Channel Model for Monostatic Sensing}
% the express of the hybrid model for the scenario
As shown in Sec.~\ref{sec:hybird_def}, the hybrid channel modeling method can be employed. Different from the ray-tracing-statistical hybrid channel model in~\cite{thz_hybrid_model} that is developed for THz communication only, we extend this hybrid model for THz monostatic sensing. In the monostatic sensing case, the hybrid channel impulse response (CIR) $h_{\rm hy}(\tau,\varphi,f)$ can be expressed as
\begin{align}
h_{\rm hy}(\tau,\varphi,f) = h_{\rm tar}(\tau,\varphi,f) + h_{\rm env}(\tau,\varphi,f),
\end{align}
where $h_{\rm tar}(\tau,\varphi,f)$ and $h_{\rm env}(\tau,\varphi,f)$ represent the target-related CIR and the environment-related CIR, respectively. Note that the target-related CIR
consists of specular reflections, while the environment-related CIR contains diffuse reflections.

%% target-related CIR
In indoor scenarios, walls are the main objects, which should be correctly characterized for monostatic sensing cases. Thus, in this work, we take the reflections from the flat wall as examples to describe the proposed hybrid channel modeling method. The target-related CIR can be generated and modeled by the target-related approach, e.g., ray-tracing, which has been discussed in~\cite{thz_hybrid_model,hybrid_isac}. It is important to investigate the power consistency of the specular reflections from the wall. 

\begin{figure}
\centering
\includegraphics[width=0.8\columnwidth]{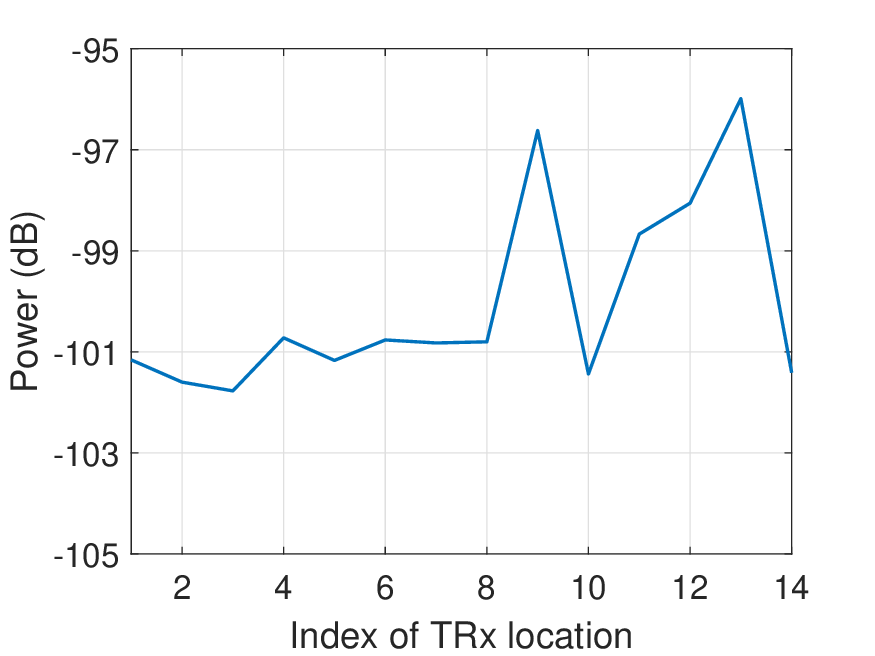}
\caption {The specular reflection power at TRx\,$1$-$14$.}\label{fig:specular_model}
\end{figure}
\begin{figure}
\centering
\includegraphics[width=0.8\columnwidth]{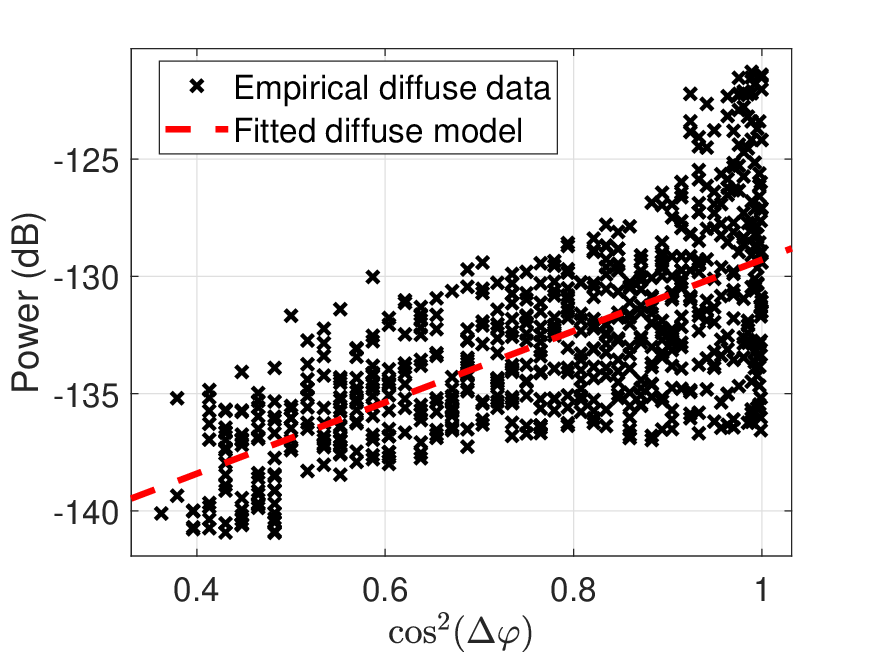}
\caption {The relationship between the diffuse power with $\cos^{2}(\Delta  \varphi)$ and the fitted diffuse power model.}\label{fig:diffuse_pow_fit}
\end{figure}

% consistency of the specular reflections from the wall
We consider the power of the specular reflected MPC at a rotation angle of 90°, representing the angle of vertical incidence and departure from the wall surface, as the specular power. Fig.~\ref{fig:specular_model} illustrates the specular power at TRx\,$1$-$14$. With the same distance from the TRx to the wall, the power consistency for monostatic sensing can be investigated. It is evident that the specular power at TRx\,$9$, $11$-$13$ is significantly higher than that from the other TRx locations. This increase in reflected power is mainly due to the reflections from the plastic sockets on the wall. By removing these TRx locations, the reflected power from the wall is observed to remain consistent, ranging from $-101.8$\,dB to $-100.7$\,dB.

%% environment-related CIR
As for the environment-related CIR, in the monostatic sensing case, the MPCs in the environment-related part are mainly contributed by the diffuse reflections. According to the classic diffuse reflection model, i.e., the Lambertian model~\cite{diffuse_model1}, the diffuse power is proportional to $\cos^{2}(\Delta \varphi)$, where $\Delta \varphi$ is the angle difference between the diffuse angle and the specular reflection angle. Thus, we propose a diffuse model $p_{\rm diff}(\Delta \varphi)$, which is related to the specular reflection power
\begin{align}
p_{\rm diff}(\Delta \varphi) = n_{\rm diff} \cdot \cos^{2}(\Delta \varphi) + b_{\rm diff},
\end{align}
where $n_{\rm diff}$ and $b_{\rm diff}$ are the slope of the diffuse power compared to $\cos^{2}(\Delta  \varphi)$ and the intercept.

\begin{figure}
\centering
    \subfigure []{\includegraphics[width=0.8\columnwidth]{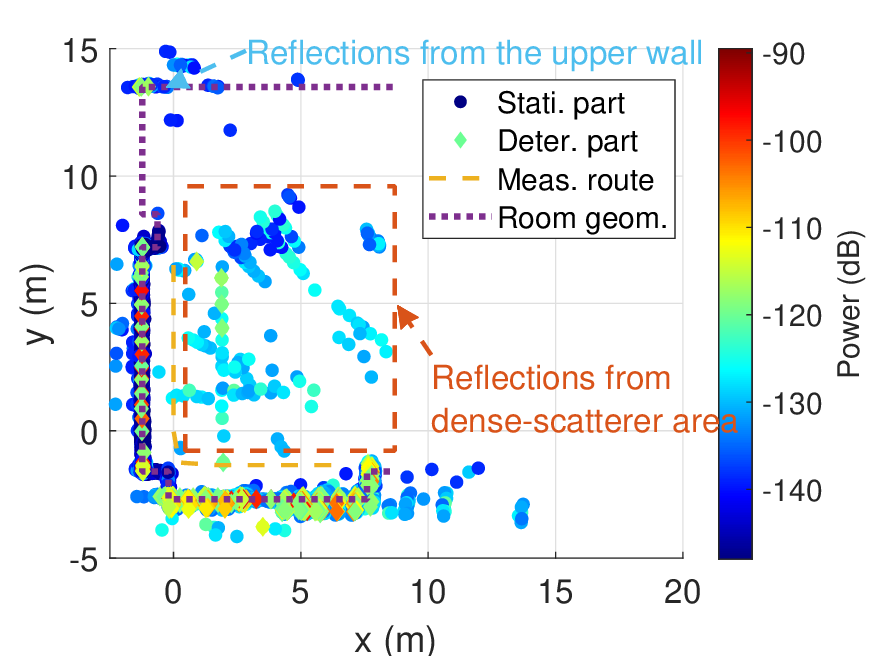}}
    \subfigure []{\includegraphics[width=0.8\columnwidth]{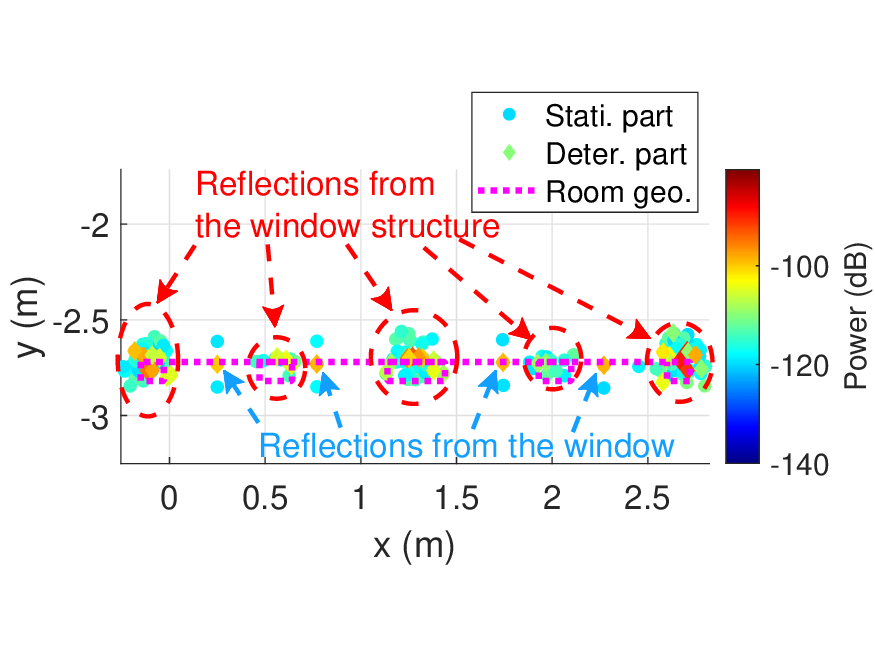}}
    \subfigure []{\includegraphics[width=0.8\columnwidth]{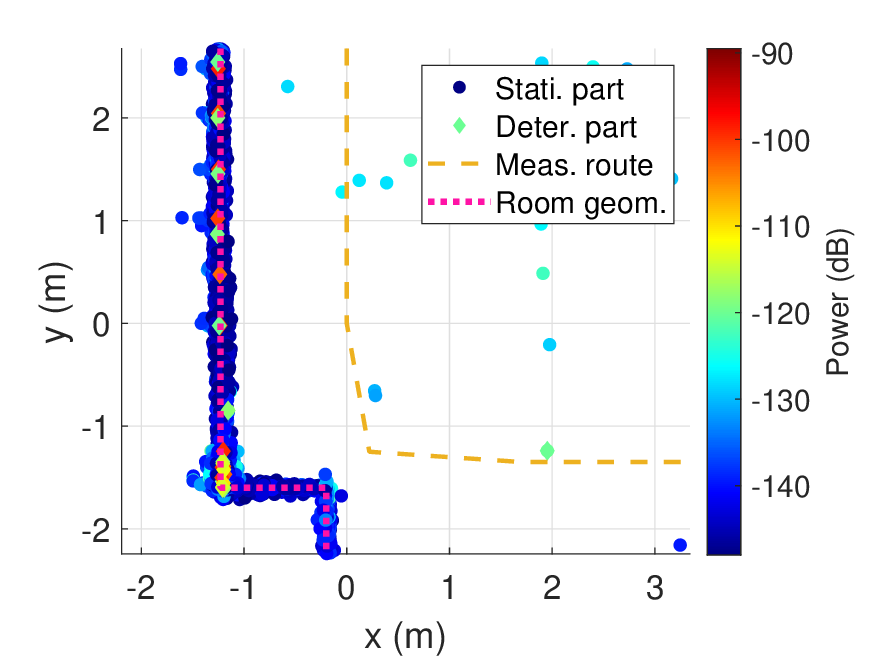}}
\caption {Demonstration of the geometric environment reconstruction. (a) The correlation between the reconstructed geometry and the $2$D room layout. (b) Zoom-in area of the left-side window in the $x$-axis range of $[-0.5,3]$\,m and the $y$-axis range of $[-3.2,-1.8]$\,m. (c) Zoom-in area of the wall in the $x$-axis range of $[-1.5,3]$\,m and the $y$-axis range of $[-2.4,1.5]$\,m.}\label{fig:room_geo}
\end{figure}

\begin{figure}
\centering
\includegraphics[width=0.8\columnwidth]{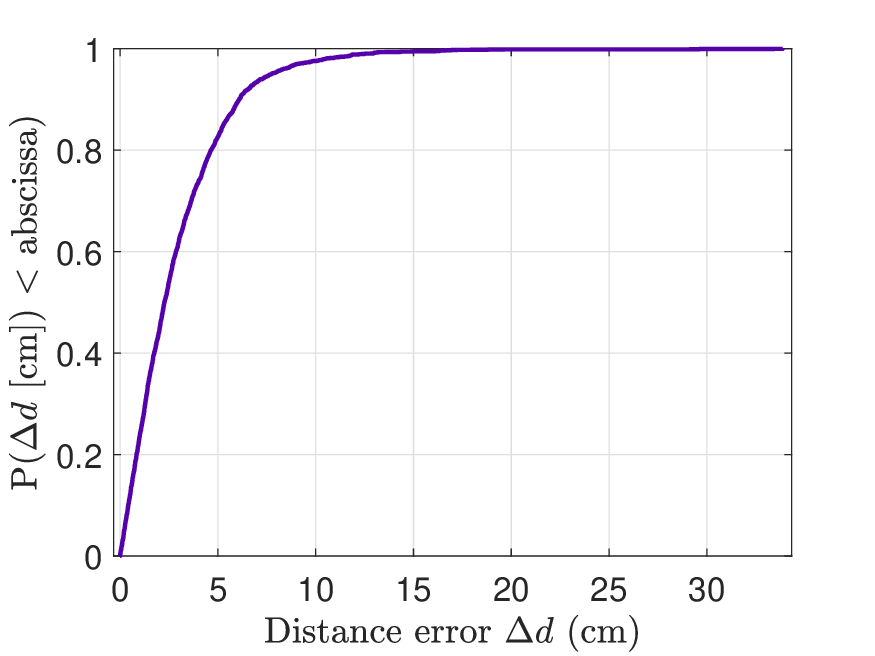}
\caption {The CDF of the distance error.}\label{fig:dis_error}
\end{figure}

% diffuse model for the wall
The relationship between the diffuse power from the flat wall and $\cos^{2}(\Delta \varphi)$ at TRx$1$-$14$ is depicted in Fig.~\ref{fig:diffuse_pow_fit}. Note that the power difference on the free space path loss is compensated. It can be observed that as $\cos^{2}(\Delta \varphi)$ increases, the diffuse power tends to increase as well, which proves the proposed diffuse model. The fitted model is also shown in Fig.~\ref{fig:diffuse_pow_fit} and compared with the empirical diffuse data. The fitted model is $p_{\rm diff}(\Delta \varphi) = 15.2 \cdot \cos^{2}(\Delta \varphi) - 144.5$ dB with the root mean square error (RMSE) of $2.95$.

\subsection{Material Identification and Environment Reconstruction}
The target-related part (i.e., specular reflection) and environment-related part (i.e., diffuse reflection) of the channel model are synthesized for environment reconstruction. First, the environment-related part can be used for the geometrical environment reconstruction. Fig.~\ref{fig:room_geo}~(a) illustrates the geometrical environment reconstruction results based on the hybrid model. Note that we use the location of TRx\,$14$ as the origin point of the coordinate, as depicted in Fig.\,\ref{fig:meas_pic}\,(b). It can be observed that the reconstruction results fit well with the room geometry. Besides, reflections from the upper wall are detected, and since the door is open, several reflections from outside the laboratory are also captured. We also zoom into the left-side window and wall areas in Fig.~\ref{fig:room_geo}~(b) and Fig.~\ref{fig:room_geo}~(c), respectively. In Fig.~\ref{fig:room_geo}~(b), the reflections from the window structures are clearly divided into five distinct groups, closely matching the window layout. Additionally, only a few strong specular reflections are observed due to the smooth surface of the window. In Fig.~\ref{fig:room_geo}~(c), the geometrical reconstruction results of the wall also match well with the wall geometry. Moreover, the corner is clearly distinguishable. The cumulative distribution function (CDF) of the distance error $\Delta d$ between the reconstruction results of the wall and the geometry is shown in Fig.~\ref{fig:dis_error}. The distance error is observed to range from $0$ to $33.8$\,cm, with an average error of $3.0$\,cm. These results demonstrate the feasibility of ultra-precise THz localization and environment reconstruction. 

\begin{figure}
\centering
    \subfigure []{\includegraphics[width=0.8\columnwidth]{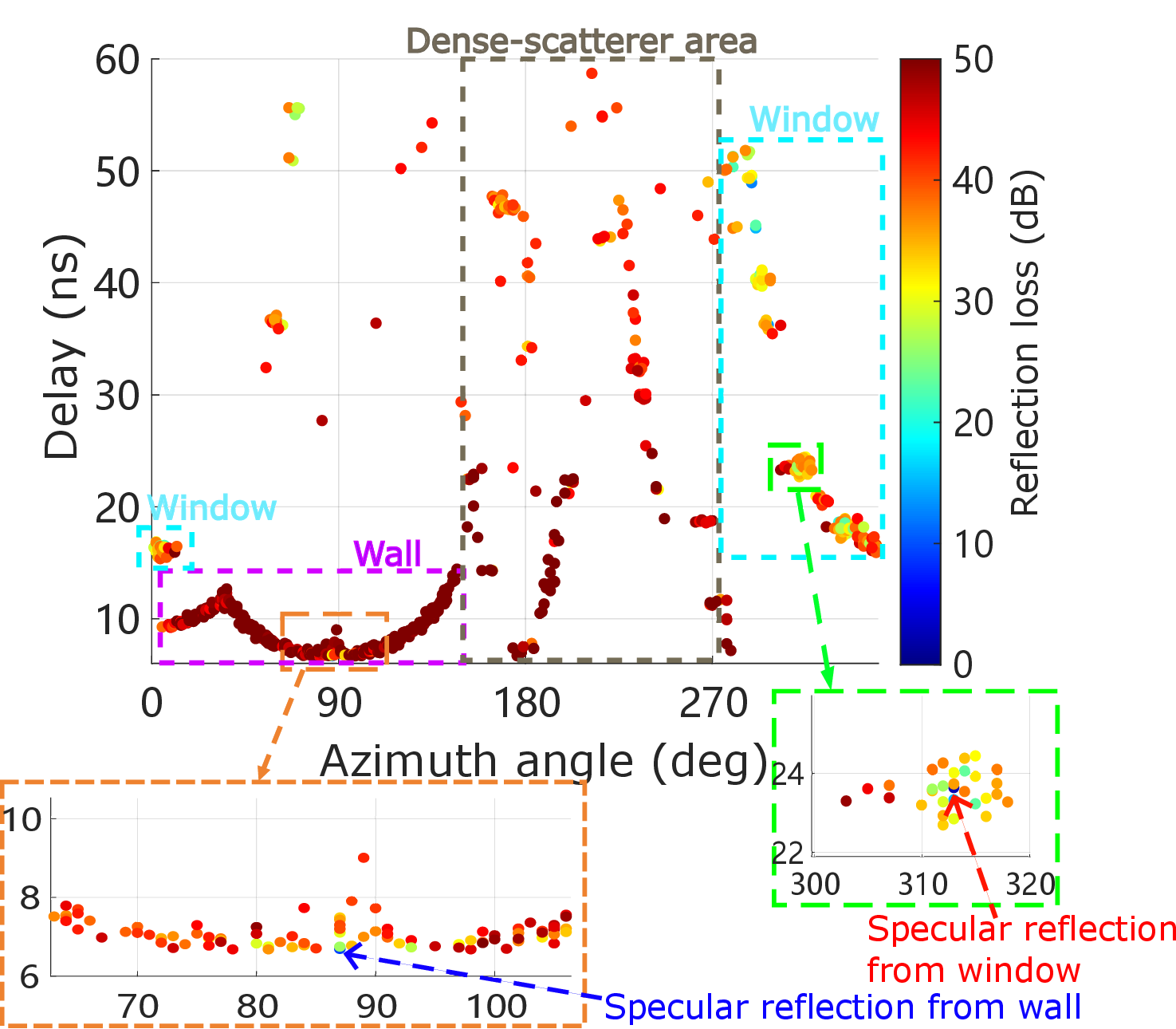}}
    \subfigure []{\includegraphics[width=0.8\columnwidth]{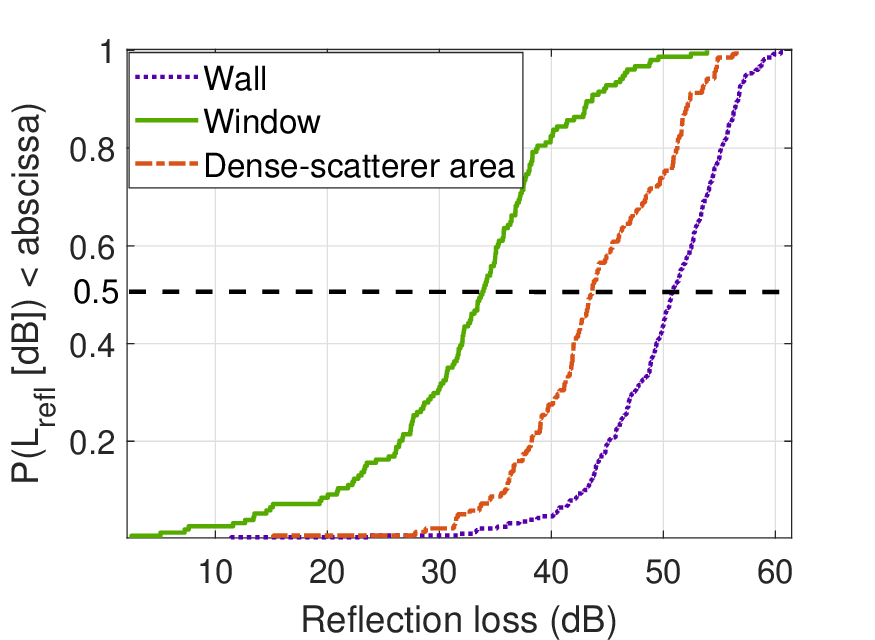}}
\caption {The exemplary reflection loss at TRx\,$14$. (a) The reflection loss in relation to the delay and azimuth angle. (b) The CDF of the reflection loss.}\label{fig:refl_loss}
\end{figure}

Reflection loss can be regarded as a parameter to detect and identify the materials in the sensing channels, which are mainly based on the target-related components from the channel. The reflection loss $L^{(m)}_{\rm refl.,\ell}$ for $\ell^{\rm th}$ de-embedded MPC at the azimuth of $\varphi$ in $m^{\rm th}$ TRx location can be expressed as
\begin{align}
L^{(m)}_{\rm refl.,\ell} [\rm dB] =  \rm FSPL^{(m)}_{\ell} [\rm dB] - 20\cdot \log_{10}|\hat{\alpha}^{(m)}_{\ell}|,
\end{align}
where $\rm FSPL^{(m)}_{\ell}$ (in decibel) is the attenuation through the propagation channel. Our SAGE estimation assesses each rotation angle independently, focusing on a single antenna configuration. Consequently, the far-field distance for this study is calculated as $d_{\rm far} = 2D^{2}/\lambda = 0.30$\,m, where $D = 12.2$\,mm is the aperture of the horn antenna used, and $\lambda$ is the wavelength corresponding to the carrier frequency of $300$\,GHz. This calculation ensures that all estimated MPCs fall within the far-field region. The attenuation $\rm FSPL^{(m)}_{\ell}$ can be calculated as
\begin{align}
{\rm FSPL^{(m)}}_{\ell} = 20\cdot \log_{10}\left( \frac{4\pi f_{c} d^{(m)}_{\ell}}{c}\right),
\end{align}
with $f_{c}$ denoting the centering frequency, i.e., $300$\,GHz, and $d^{(m)}_{\ell} = \hat{\tau}^{(m)}_{\ell}\cdot c$ representing the propagation distance of $\ell^{\rm th}$ path. 

The exemplary reflection loss with the delay and azimuth angle at TRx\,$14$ is shown in Fig.~\ref{fig:refl_loss}~(a). The reflection losses vary from $2.5$\,dB to $60.7$\,dB. Besides, the specular reflection losses from the wall and the metallic window structure are $11.4$\,dB and $2.5$\,dB, respectively. Based on the geometry, we can simply separate the reflection losses into three groups, i.e., wall, window, and dense-scatterer areas, as depicted in Fig.~\ref{fig:refl_loss}~(b). The CDF of the reflection loss from all the TRx locations is illustrated in Fig.~\ref{fig:refl_loss}~(b). The reflection losses from the window range between $2.5$\,dB and $54$\,dB, while those from the dense-scatterer area are slightly higher, between $15.3$\,dB and $56.6$\,dB. The wall reflections experience the most significant losses, ranging from $11.4$\,dB to $60.7$\,dB. The metallic structure of the window leads to comparatively lower reflection losses, whereas the cement wall reflects less power and its rough surface causes more diffuse reflections, resulting in higher losses. The average reflection losses are $33.7$\,dB for the window, $43.5$\,dB for the dense-scatterer area, and $50.8$\,dB for the wall. These results highlight the need for detailed material measurements and identification at THz frequencies.

\subsection{Delay and Angular Dispersion}
\subsubsection{Delay spread}
As for the delay dispersion, the root-mean-square (RMS) delay spread $\sigma _{\tau}$ is a channel parameter to describe the delay span of the MPCs in the channel~\cite{lyu_tracking,del_spread}. For the monostatic sensing case, each rotation angle contains MPC information for positioning and sensing. Therefore, in this work, the delay spread is calculated based on the de-embedded MPC results, which is a function of the rotation angle.

\begin{figure}
\centering
    \subfigure []{\includegraphics[width=0.8\columnwidth]{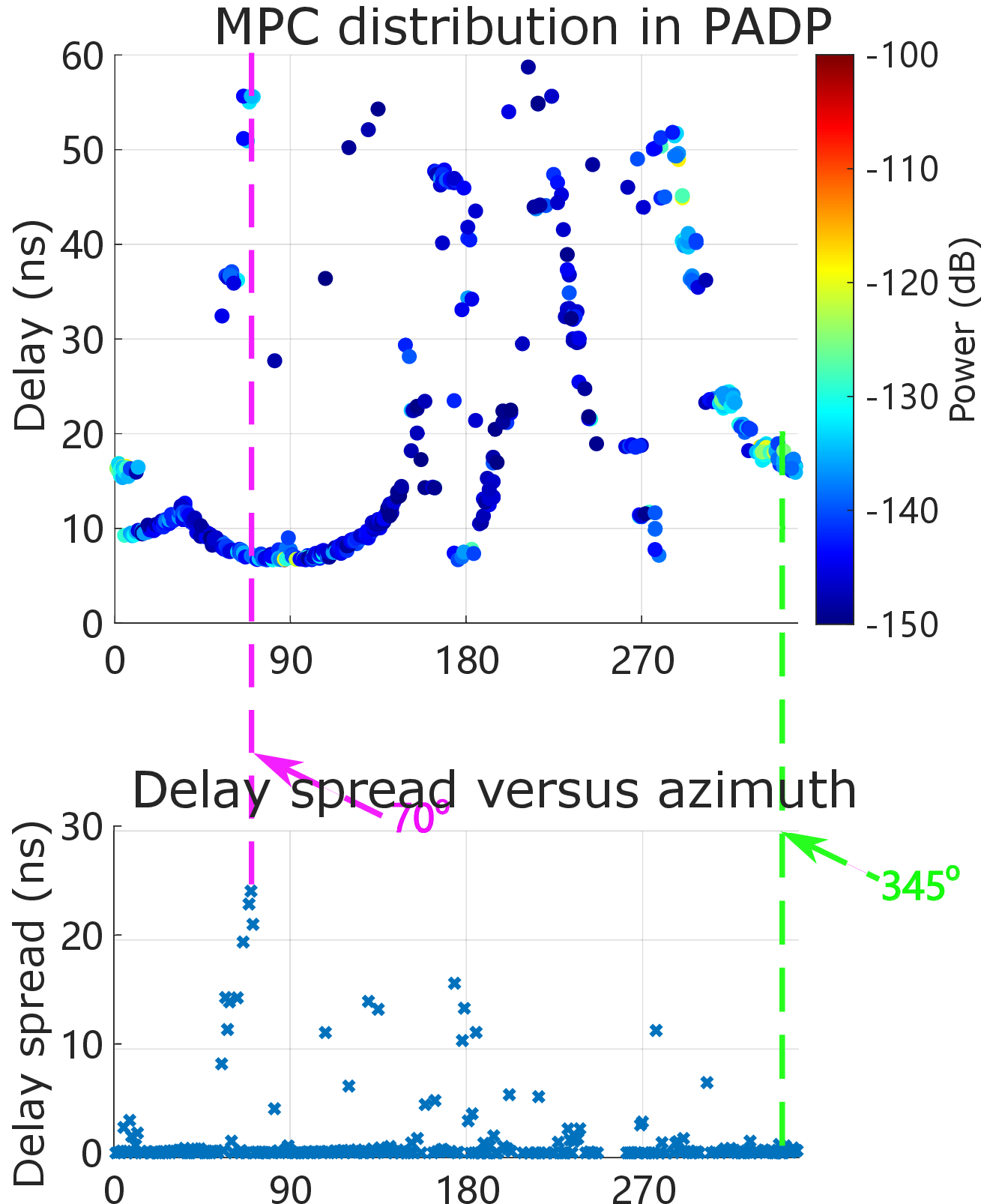}}
    \subfigure []{\includegraphics[width=0.8\columnwidth]{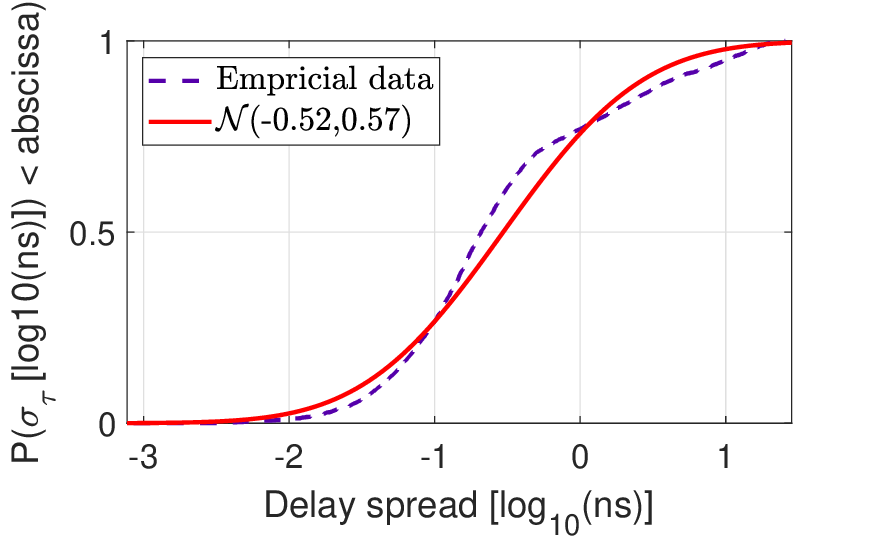}}
\caption {The delay spread results. (a) Exemplary delay spreads versus rotation angle at TRx\,$14$ (bottom subfigure) compared with the estimated MPC distribution (upper subfigure). (b) The CDF of the empirical delay spread and the fitted model.}\label{fig:tau_spread}
\end{figure}

The comparison between the exemplary delay spread versus azimuth angle at TRx\,$14$ and the MPC distribution in PADP is depicted in Fig.~\ref{fig:tau_spread}~(a). The delay spread varies in the range of $[0,24.1]$\,ns. It can be observed that the majority of the delay spreads ($242$ out of $360$ azimuth) are relatively low, concentrated within the delay range of $[0,5]$\,ns, for example, at the angle of $345^{\circ}$, the delay spread exhibits a low value of $0.7$\,ns since the MPCs at this angle is concentrated in the delay range of $[16.5, 18.1]$\,ns. Moreover, some certain azimuth angles ($85$ out of $360$ azimuth angles) show a delay spread of $0$\,ns, indicating that only a single MPC is detected at those angles. At this TRx position, several high delay spreads are observed. For instance, at an azimuth of $70^{\circ}$, a delay spread of $24.1$\,ns is observed. At this angle, a prominent MPC with a long delay of $55.6$\,ns is noticeable, contributing to the increased delay spread. In Fig.~\ref{fig:tau_spread}~(b), the CDF of the calculated delay spread (shown on a logarithmic scale) is presented, along with the corresponding fitted results. Note that the delay spreads with a value of $0$\,ns have been filtered out for CDF analysis. Notably, it is observed that $76$\% delay spreads are lower than $1$\,ns. The CDF of the logarithmic delay spread shows a close fit to a normal distribution with parameters $\mathcal{N}(-0.52,0.57)$. 

\subsubsection{Angular spread}
\begin{figure}
\centering
    \subfigure []{\includegraphics[width=0.8\columnwidth]{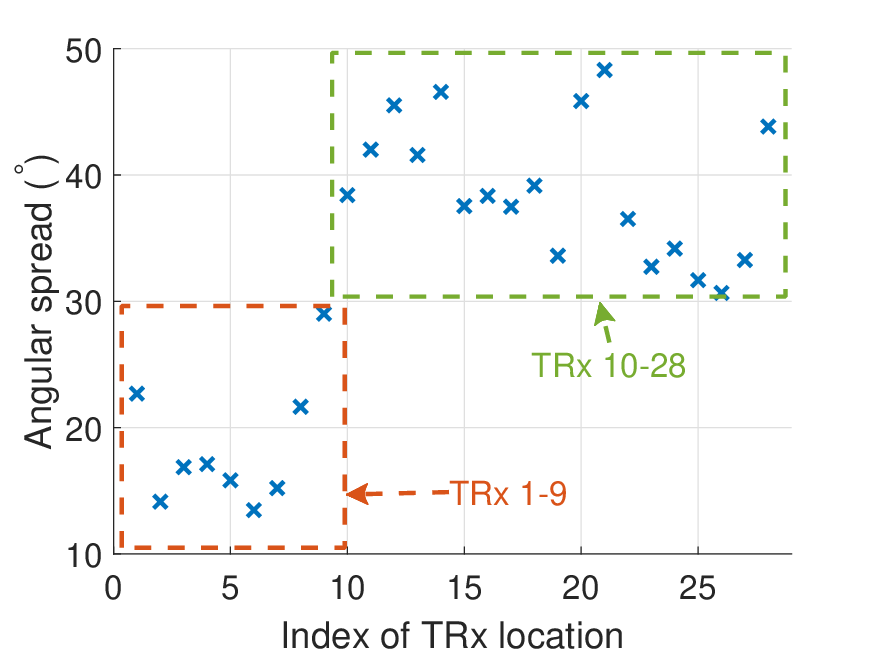}}
    \subfigure []{\includegraphics[width=0.8\columnwidth]{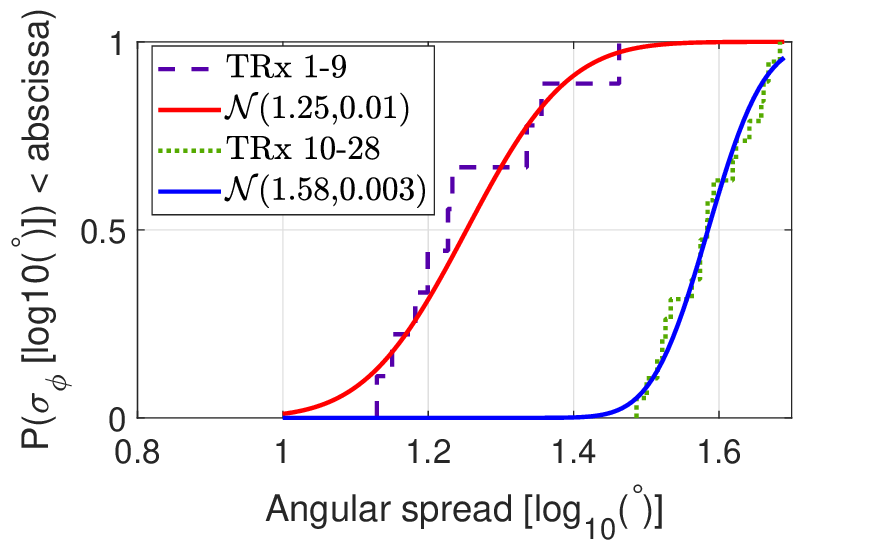}}
\caption {The angular spread results. (a) The relationship between the angular spread and the TRx locations. (b) The CDF of the empirical angular spread and the fitted model.}\label{fig:ang_spread}
\end{figure}

The angular spread $\sigma _{\phi}$ is a parameter to characterize the angle dispersion, which can be referred to in~\cite{aoa_spread,aoa_spread2,3GPP_38.901}. The relation with the angular spread to the TRx locations is shown in Fig.~\ref{fig:ang_spread}~(a). The angular spreads at TRx $1$-$8$ are relatively low, ranging from $[13.5^{\circ},22.7^{\circ}]$. However, a significant increase in angular spreads from $15.2^{\circ}$ to $42.0^{\circ}$ is observed from TRx\,$7$ to TRx\,$11$. This is due to the shorter distance between the TRx and the window, as well as strong reflections from the metallic structure of the window, which contribute to the substantial angular spread. From TRx\,$11$ to TRx\,$28$, the angular spreads are noticeably higher, ranging from $30.7^{\circ}$ to $48.3^{\circ}$. This increase can be attributed to the wide separation between the two windows and the strong reflections from the metallic structure of the windows. From Fig.~\ref{fig:ang_spread}~(a), it is evident that the angular spread values for TRx\,$1$-$9$ and TRx\,$10$-$28$ follow two distinct distributions. Fig.~\ref{fig:ang_spread}~(b) shows the CDF of the angular spread on a logarithmic scale, along with the fitted angular spread model. The empirical logarithmic angular spread for TRx\,$1$-$9$ and TRx\,$10$-$28$ closely aligns with normal distributions $\mathcal{N}(1.25, 0.01)$ and $\mathcal{N}(1.58, 0.003)$, respectively.

\section{Conclusion}\label{sec:conclusion}
In this work, channel measurements were carried out in an indoor laboratory scenario exploring the monostatic-sensing channels in the frequency range of $290$-$310$\,GHz. The Tx and Rx are placed in the same location. The DSS scheme is employed to obtain the spatial sensing-channel profiles in the azimuth range from $0^{\circ}$ to $359^{\circ}$ with a rotation step of $1^{\circ}$. With such high delay and angular resolution, the MPCs can be clearly distinguished in both delay and azimuth domains. The measurements take $28$ TRx locations in the scenario with a 'L'-shaped route. An HRPE algorithm, i.e., element-wise SAGE algorithm, is first used to estimate the amplitudes and delays of the MPCs. The specular and diffuse reflections are then analyzed based on the geometry. A geometry-based MPC trajectory tracking algorithm is proposed to track the MPCs with the same radiation pattern effects and de-embed the radiation pattern. After obtaining the de-embedded MPCs, a hybrid channel model is proposed. Besides, the channel characteristics, i.e., reflection loss, delay spread, and angular spread, are analyzed based on the de-embedded MPCs. This work offers valuable insights into THz monostatic sensing channel modeling and the design of future THz ISAC systems. Future efforts will focus on developing a comprehensive hybrid ISAC channel model for various indoor and outdoor scenarios, as well as on THz material identification.

%\section*{Acknowledgement}
%The authors would like to thank Yifa Li, Sigurd S$\mathrm{\acute{a}}$ndor Petersen, and Kim Olesen for their help with the measurements. The work was partially supported by the $21$NRM$03$ MEWS project, which has received funding from the European Partnership on Metrology, co-financed from the European Union’s Horizon Europe Research and Innovation Programme, and by the Participating States, and also supported by the Grant named Short-Term Scientific Mission (STSM) from COST CA$20120$ INTERACT.

\normalem
\bibliography{monostatic_sensing}

\end{document}